%% file: JCGS_ARXIV.tex
\documentclass[12pt]{article}
\usepackage{amsthm,amsmath,amsfonts,amssymb}
\usepackage{mathdef}
\usepackage{natbib}

\usepackage[pdf]{pstricks}
\usepackage{pst-node,pst-plot}
\usepackage{graphicx,pst-all}
\usepackage{enumerate}
\usepackage{natbib}
\bibpunct{(}{)}{,}{a}{}{,}
\ifx\bibfont\newcommand{\bibfont}{\small}
\else\renewcommand{\bibfont}{\small}\fi
\newcommand*\tageq{\refstepcounter{equation}\tag{\theequation}}
\graphicspath{{GraphsMay2018/Monkey/}{GraphsMay2018/wnbreak/}{GraphsMay2018/mabreak/}{GraphsMay2018/slices/}{GraphsMay2018/ma2/}}
\usepackage[utf8]{inputenc}
\usepackage{setspace}
\usepackage{graphicx}
\usepackage{multirow}
\usepackage{lscape}
\usepackage{afterpage}
\usepackage{xcolor}
\usepackage{subcaption}
\usepackage[section]{placeins}
\usepackage{flafter}
\usepackage{verbatim}
\usepackage{mathrsfs}
\usepackage{bbm}
\usepackage{url}
\usepackage{todonotes}

\makeatletter
\renewcommand{\todo}[2][]{%
    \@todo[caption={#2}, #1]{\begin{spacing}{0.5}#2\end{spacing}}%
} 
\makeatother 

\newcommand{\blind}{0}
\newcommand{\hatfT}{\hat f_{T}}
\newcommand{\hatfkT}{\hat f^{(k)}}
\newcommand{\tildefkT}{\tilde f^{(k)}}
\newcommand{\tildefrelkT}{\tilde f^{(k)}_{\text{rel}}}
\newcommand{\barfkT}{\bar f^{(k)}}
\newcommand{\barfrelkT}{\bar f^{(k)}_{\text{rel}}}
\newcommand{\hatfkkT}[1]{\hat f^{(#1)}}
\newcommand{\tildefkkT}[1]{\tilde f^{(#1)}}
\newcommand{\Kf}{K_{\text{f}}}
\newcommand{\Kt}{K_{\text{t}}}
\newcommand{\alphaf}{\alpha_{\text{f}}}
\newcommand{\alphat}{\alpha_{\text{t}}}
\newcommand{\Kst}{K_{\text{p}}}

\newcommand{\kappaf}{\kappa_{\text{f}}}
\newcommand{\kappat}{\kappa_{\text{t}}}

\newcommand{\baf}{b_{\text{f}}}
\newcommand{\bt}{b_{\text{t}}}
\newcommand{\bfT}{b_{\text{f},T}}
\newcommand{\btT}{b_{\text{t},T}}

\newcommand{\bfkT}{b_{\text{f},T}^{(k)}}
\newcommand{\btkT}{b_{\text{t},T}^{(k)}}
\newcommand{\bfkkT}[1]{b_{\text{f},T}^{(#1)}}
\newcommand{\btkkT}[1]{b_{\text{t},T}^{(#1)}}

\newcommand{\JT}{J_T}
\newcommand{\NkT}{N^{(k)}}
\newcommand{\MkT}{M^{(k)}}
\newcommand{\tildeNkT}{\tilde N^{(k)}}
\newcommand{\hatNkT}{N^{(k)}}
\newcommand{\tildeMkT}{\tilde M^{(k)}}
\newcommand{\tildeMkkT}[1]{\tilde M^{(#1)}}
\newcommand{\hatMkT}{M^{(k)}}
\newcommand{\tildeNkkT}[1]{\tilde N^{(#1)}}

\newcommand{\phiT}{\phi^{(T)}_{u,\lam}}

\newcommand{\tildewk}[1]{\tilde w^{(k)}_{u,\lam}}
\newcommand{\barwk}[1]{\bar w^{(k)}_{u,\lam}}

\newcommand{\WkT}{W^{(k)}}
\newcommand{\tildeWkT}{\tilde W^{(k)}}

\newcommand{\tildesigkT}{\tilde\sigma^{(k)}_{T}}
\newcommand{\fmax}{\hat{f}^{(k_{\text{{final}}})}_{\text{a}}}

\newcommand{\fnaopt}{\hat{f}^{b_{\text{opt}}}_{\text{na}}}
\newcommand{\fna}{\hat{f}_{\text{na}}}

\newcounter{alphcount}
\newenvironment{alphlist}%
{\begin{list}{{\upshape(}\alph{alphcount}\/{\upshape)\ }}%
             {\usecounter{alphcount}\labelwidth1.5em%
              \leftmargin2em\labelsep0.5em\topsep0.25em plus 0.5ex%
              \itemsep0.25em plus 0.5ex\parsep0em}}{\end{list}}
{\begin{list}{{\upshape(#1\arabic{alphcount})\hfill}}%
             {\usecounter{alphcount}\labelwidth2.5em%
              \leftmargin2.5em\labelsep0em\topsep0.25em plus 0.5ex%
              \itemsep0.25em plus 0.5ex\parsep0em}}{\end{list}}
{\begin{list}{{\upshape\arabic{alphcount}.\ }}%
             {\usecounter{alphcount}\labelwidth1.5em%
              \leftmargin2em\labelsep0.5em\topsep0.25em plus 0.5ex%
              \itemsep0.25em plus 0.5ex\parsep0em}}{\end{list}}
\newcounter{romancount}
\newenvironment{romanlist}%
{\begin{list}{{\upshape(}\roman{romancount}\/{\upshape)\ }}%
             {\usecounter{romancount}\labelwidth2em%
              \leftmargin2.5em\labelsep0em\topsep0.25em plus 0.5ex%
              \itemsep0.25em plus 0.5ex\parsep0em}}{\end{list}}
{\begin{list}{$\bullet$\ }%
             {\labelwidth1.5em%
              \leftmargin2em\labelsep0.5em\topsep0.25em plus 0.5ex%
              \itemsep0.25em plus 0.5ex\parsep0em}}{\end{list}}
\newcounter{storecount}
\newcommand{\savecounter}[1]{\setcounter{storecount}{\value{#1}}}
\newcommand{\restorecounter}[1]{\setcounter{#1}{\value{storecount}}}

\begin{document}

\def\spacingset#1{\renewcommand{\baselinestretch}%
{#1}\small\normalsize} \spacingset{1}


\if0\blind
{
  \title{\bf Data-adaptive estimation of time-varying \\ spectral densities}
  \author{Anne van Delft\thanks{ This work has been supported in part by Maastricht University, the contract ``Projet d'Actions de Recherche Concert{\'e}es'' No. 12/17-045 of the ``Communaut{\'e} fran\c{c}aise de Belgique'' and by the Collaborative Research Center ``Statistical modeling of nonlinear dynamic processes'' (SFB 823, Project A1, A7) of the German Research Foundation (DFG). 
  } \hspace{.2cm}\\
  Fakult{\"a}t f{\"u}r Mathematik, Ruhr-Universit{\"a}t Bochum\\
   {\em E-mail address:} anne.vandelft@rub.de\\
    and \\
    Michael Eichler\\
    Department of Quantitative Economics,
Maastricht University, \\
{\em E-mail address:} m.eichler@maastrichtuniversity.nl}
  \maketitle
} \fi

\if1\blind
{
  \bigskip
  \bigskip
  \bigskip
  \begin{center}
    {\LARGE\bf Title}
\end{center}
  \medskip
} \fi

\bigskip
\begin{abstract}
This paper introduces a data-adaptive non-parametric approach for the estimation of time-varying spectral densities from nonstationary time series. Time-varying spectral densities are commonly estimated by local kernel smoothing. The performance of these nonparametric estimators, however, depends crucially on the smoothing bandwidths that need to be specified in both time and frequency direction. As an alternative and extension to traditional bandwidth selection methods, we propose an iterative algorithm for constructing localized smoothing kernels data-adaptively. The main idea, inspired by the concept of propagation-separation \citep{Polzehl2006}, is to determine for a point in the time-frequency plane the largest local vicinity over which smoothing is justified by the data. By shaping the smoothing kernels nonparametrically, our method not only avoids the problem of bandwidth selection in the strict sense but also becomes more flexible. It not only adapts to changing curvature in smoothly varying spectra but also adjusts for structural breaks in the time-varying spectrum. 
\end{abstract}

\noindent%
\vfill

\newpage
\spacingset{1.45} 
\input{Introduction.tex}
\input{Sec2-localstat.tex}

\input{Sec3a-propag.tex}
\input{Sec3b-details.tex}
\input{Sec5-examp.tex}
\input{Sec6-appl.tex}
\input{Sec7-disc.tex}

\bigskip
\begin{center}
{\large\bf SUPPLEMENTARY MATERIAL}
\end{center}

\begin{description}\label{supp}
\item[{\bf Supplement: }] Supplement with additional details on the distributional properties of the adaptive kernel estimator \citep{vDE18supp}.
\item[{\bfseries R-code:} ]
Source code of an implementation of the described algorithm in {\tt R} and {\tt C++}; instructions for running the code are provided in the file {\em readme.pdf} in the root folder. ({\em RoutineDataAdapt.zip}, zip file)
\end{description}
\bibliographystyle{stat}

\newpage
\setcounter{section}{0}
\setcounter{equation}{0}
\def\theequation{\arabic{section}.\arabic{equation}}
\def\thesection{\arabic{section}}
\setcounter{footnote}{0}

\begin{center}
 {\LARGE \bf { Supplement to \\ ``Data-adaptive estimation of time-varying spectral densities''\footnote{
 This work has been supported in part by Maastricht University, the contract ``Projet d'Actions de Recherche Concert{\'e}es'' No. 12/17-045 of the ``Communaut{\'e} fran\c{c}aise de Belgique'' and by the Collaborative Research Center ``Statistical modeling of nonlinear dynamic processes'' (SFB 823, Project A1, A7) of the German Research Foundation (DFG).} }}
 \vspace{10pt}
 \hspace{5cm}\\
{\large Anne van Delft\\
  Fakult{\"a}t f{\"u}r Mathematik, Ruhr-Universit{\"a}t Bochum\\
   {\em E-mail address:} anne.vandelft@rub.de\\
    and \\
    Michael Eichler\\
    Department of Quantitative Economics,
Maastricht University, \\
{\em E-mail address:} m.eichler@maastrichtuniversity.nl\\}
  \maketitle
\end{center}
\bigskip
\begin{abstract}
This supplement contains technical details on the distributional properties of the adaptive estimator as defined in the main paper \citep{vDE18}. 
Using the setting of empirical spectral processes \citep[e.g.][]{Dahlhaus2009b,Dahlhaus2009a}, we conclude that asymptotic properties of the nonadaptive estimator carry over to the adaptive estimator under local homogeneity.
\end{abstract}

\noindent%
\vfill

\newpage
\spacingset{1.45} 
\input{Sec3c-asympt.tex}
\bigskip

\bibliographystyle{stat}

\end{document}

%% file: Introduction.tex
\section{Introduction}

Spectral analysis of time series data has been of interest for many years and has a varied history owing to applications in a wide range of disciplines such as geophysics, astronomy, sound analysis, analysis of medical data or yet of economical data. There exists a rather extensive literature on spectral analysis of weakly stationary processes and statistical techniques are well developed \citep{Cramer1942, Bartlett1950, Grenander1957, Cooley, Brillinger}. However, in many applications the time series at hand show some nonstationary behavior in the sense that the oscillations described by the spectrum change over time. For instance, such behavior can be observed for the brain activity in various brain regions during associative learning experiments \citep{Fiecas2014}. In such cases, the assumption of weak stationarity often still seems plausible over shorter time periods, that is, the series can locally be well approximated by a stationary time series and its local oscillations can be described by the spectrum of the approximating stationary series. Empirically, this idea is reflected, for example, in the use of the spectrogram or the segmented periodogram for time-frequency data analysis \citep[e.g.][]{Sandsten2016}.

Theoretically, several definitions of time-varying spectra have been proposed in the literature \citep[e.g.][]{Priestley1965,Subba1970,Martin1985,Hallin1986}. Most definitions depend on the length $T$ of the time series and hence are problematic for statistical inference. For a unifying approach, \citet{Dahlhaus1996a} developed the concept of locally stationary processes, where the nonstationary time series is embedded in a sequence of series that share the same dynamics over an increasing number of observations. This framework not only yields a unique definition of a time-varying spectrum but also allows meaningful asymptotic approximations to the sampling distribution of localized estimators and test statistics.

Like in the stationary case, time-varying spectral densities are commonly estimated by kernel smoothing of some raw spectral estimator such as the segmented periodogram or the pre-periodogram \citep{Neumann1997}. An important problem that is common to all kernel smoothing estimators is the selection of proper bandwidths since the accuracy of the resulting estimators has been found to be quite sensitive to the choice of bandwidths \citep[e.g.][]{Eichler2011}. Theoretical optimal bandwidths \citep{Dahlhaus1996a} depend on the unknown underlying spectral density, and no guidelines are available on how to set them in practice  \citep{Dahlhaus2009a}. To our knowledge, data-adaptive schemes suitable in the context of time-frequency analysis have not yet been considered. However, under specific parametric assumptions there are methods available based on the segmented periodogram. For example, \citet{Sergides2009} and \citet{Preu2011} use an integrated version of the segmented periodogram to test for semi-parametric hypotheses, which avoids selection of  bandwidths in frequency direction. The major drawback of using the segmented periodogram as an underlying estimator is that a fixed bandwidth in time direction, namely the length of segments, must be set. This leads to seriously biased estimates in case of structural breaks in the spectrum.

In this paper, we propose an alternative approach that circumvents the problem of classical bandwidth selection by data-adaptive shaping of the smoothing kernel. The procedure is inspired by the propagation-separation approach, an adaptive weighing scheme where the iteratively updated weights gain from previously aggregate information. This method was first introduced by \citet{Polzehl2006} in the context of local likelihood models and is closely related to previous works in this direction \citep[e.g.,][]{Lepskii1990,Polzehl2000}. The general principle is to determine, based on some measure of homogeneity, for each design point the maximal local neighborhood that can be used for smoothing. Starting with a small initial neighborhood, the smoothing region is iteratively extended to include new data points for which the hypothesis of homogeneity can still be maintained (propagation) while data points are excluded whenever it is violated. This approach has demonstrated to be useful in a variety of problems such as image denoising and classification \citep[e.g.,][]{TPSpok2008,Belspok2007}. 

In our adaptation of the propagation--separation approach, at each iteration of the algorithm the neighborhood used for estimating the time-varying spectral density $f(u, \lambda)$ at rescaled time $u\in[0,1]$ and frequency $\lambda\in[-\pi,\pi]$ is described by weights $W_{u,\lambda}(v,\mu)$ that define the shape of the kernel. These weights are derived from the spectral estimates constructed at the previous step. The effective neighborhood for local smoothing is then given by the points for which the corresponding weights are non-zero. As a discrepany measure for the deviation from homogeneity, we use the squared relative difference between the corresponding spectral estimates.
Compared to classical kernel estimates with either global or local bandwidth, the advantage of our approach is that  the smoothing kernel itself is adjusted in terms of shape and effective bandwidth for each point separately. This flexibility in shaping the smoothing kernel completely data-adaptively is of particular importance when structural breaks are present. In that case, our estimator can gain precision from smoothing in one direction without getting severely biased by smoothing across the break. The algorithm has been additionally robustified to be less affected by so-called cross-terms, which pose a serious problem in high-resolution time-frequency analysis. We emphasize that our work is related to aforementioned existing literature on data-adaptive schemes but that these are not directly applicable in the setting of time-frequency analysis.

Our paper is organized as follows. In section 2, we provide the background on locally stationary processes and estimation of time-varying spectral densities and briefly describe the notion of cross-terms. In section 3, we present the algorithm and explain the importance of the various steps. In section 4, we illustrate the properties of the proposed estimator by three examples and examine its performance in a simulation study. Finally, the approach is illustrated by an application to local field potential (LFP) recordings in section 5. The routine and further technical details are provided as supplementary material.

%% file: Sec2-localstat.tex
\section{Locally stationary processes}

Let $X_t$ be a non-stationary process that has been observed at times $t=1,\ldots,T$. For frequency-domain based analysis of the process, we follow the approach by \citet{Dahlhaus1996a} and view the process $X_t$ as part of a sequence of processes $\{X_{t,T},t=1,\ldots,T\}$, $t\in\nnum$, where $X_{t,T}$ has the representation
\[
X_{t,T}=\lsum_{j\in\znum} a_{t,T}(j)\,\veps_{t-j}
\]
for some weakly stationary white noise process $\{\veps_t\}$ with $\mean(\veps_t)=0$ and $\mean(\veps_t^2)=1$. The processes $X_{t,T}$ for different $T$ are related by (approximately) sharing the same dynamics locally. More precisely, we assume that there exists functions $a(u,\lam)$ on $[0,1]\times[-\pi,\pi]$ such that
\begin{romanlist}
\item
$\DS\sup_{t,T}|a_{t,T}(j)|\leq K\,\ell(j)^{-1}$ for all $j\in\znum$,
\item
$\DS\sup_{u}|a(u,j)|\leq K\,\ell(j)^{-1}$ for all $j\in\znum$,
\item
$\DS\sup_{j}\lsum_{t=1}^{T}\Big|a_{t,T}\big(j\big)
-a\big(\tfrac{t}{T},j\big)\Big|\leq K$,
\item
$\DS V\big(a(\cdot,j)\big)\leq K\,\ell(j)^{-1}$ for all $j\in\znum$,
\end{romanlist}
where
\[
V(g)=\sup\bigg\{\lsum_{k=1}^{m}\big|g(x_k)-g(x_{k-1})\big|\,\bigg|
\,0\leq x_0<\ldots<x_m\leq 1,\,m\in\nnum\bigg\}
\]
is the total variation of a function $g$ on $[0,1]$ and
\[
\ell(j)=\max\big\{1,|j|\,\log^{1+\zeta}|j|\big\}
\]
for some constant $\zeta>0$. A sequence of processes $\{X_{t,T}\}$ satisfying the above assumptions (i) to (iv) is called {\em locally stationary} \citep[e.g.][]{Dahlhaus2009b,Dahlhaus2009a}.

The above representation implies that, locally about a point $u\,T$ for some $u\in[0,1]$, the processes $\{X_{t,T}\}$ can be approximated by the weakly stationary process
\[
X^{(u)}_t=\lsum_{j\in\znum} a(u,j)\,\veps_{t-j},\qquad t\in\znum,
\]
and thus their the oscillating behavior of the processes $\{X_{t,T}\}$ can be
described by the spectral density
\begin{equation}
\label{eq:tvsd-def}
f(u,\lambda)=\SSS{\frac{1}{2\pi}}\,A(u,\lambda)\,A(u,\lambda)^*,
\end{equation}
where
\[
A(u,\lam)=\lsum_{j\in\znum} a(u,j)\,e^{-\im\,\lam\,j}.
\]
The function $f(u,\lambda)$ in \eqref{eq:tvsd-def} is called the time--varying spectral density of the locally stationary process $\{X_{t,T}\}$ at frequency $\lam\in[-\pi,\pi]$ and rescaled time $u\in[0,1]$.

In this paper, we are more generally interested in the estimation of the time-varying spectral densities of processes $\{X_{t,T}\}$ that possibly exhibit structural breaks in time. Therefore, we impose the following condition om the process $\{X_{t,T}\}$.
\begin{assumption}\label{smoothness} 
$\{X_{t,T},\,t = 1,\ldots,T,\,
T\in\nnum\}$ is a piecewise locally stationary process with time-varying spectral density $f(u,\lambda)$ that is twice differentiable in $\lambda$ and piecewise twice differentiable in $u$ with bounded derivatives in both directions.
\end{assumption}

Non-parametric approaches for the estimation of the time-varying spectral density are based on the fact that
\[
\gamma(u,k)=\int_{-\pi}^{\pi}f(u,\lambda)\,e^{\im\,k\,\lambda}\,d\lambda
\]
defines a localized auto-covariance function. Using estimators $\hat\gamma_T(u,k)$ for $\gamma(u,k)$, we obtain
\[
\JT(u,\lam)=\tfrac{1}{2\pi}\,\lsum_{k\in\znum}\hat\gamma_T(u,k)\,e^{-\im\,k\,\lambda}
\]
as an estimator for the time-varying spectral density at rescaled time $u=\tfrac{t}{T}$ and frequency $\lambda$. For instance, taking auto-covariance estimators on segments of length $2m$, 
\[
\hat\gamma_T\big(\tfrac{t}{T},k\big)=\SSS{\frac{1}{2m}}\,\sum_{j=t-m}^{t+m-k}
X_j\,X_{j+k}
\]
for $k\geq 0$ and $\hat\gamma_T\big(\tfrac{t}{T},k\big) = \hat\gamma_T\big(\tfrac{t}{T},-k\big)$ for $k<0$, we obtain the segmented periodogram \citep[e.g.][]{Dahlhaus1996a,Preu2011}, which treats the series over a segment of length $2m$ about the time point $u\,T$ as stationary. While this approach inherits the good properties of the periodogram in the stationary case, the restriction to lags at most $2m$ leads to a loss in the frequency resolution and the averaging over the segment to a loss in time-resolution.

For the purpose of this paper, the pre-periodogram introduced by \citet{Neumann1997} is more suitable. In its symmetrized version \citep{Jeong1992}, it is obtained from the local auto-covariance estimator
\begin{equation}\label{eq:ppgram}
\hat\gamma_T\big(\tfrac{t}{T},k\big)=
\begin{cases} 
X_{t-\frac{k}{2}}X_{t+\frac{k}{2}} & \text{if }t\pm\frac{k}{2}\in\znum,\\
\tfrac{1}{2}\big(X_{t-\frac{k-1}{2}}X_{t+\frac{k-1}{2}} +  X_{t+\frac{k+1}{2}}X_{t+\frac{k+1}{2}}\big) & \text{otherwise}.
\end{cases}
\end{equation}
The pre-periodogram exhibits a much better time-frequency concentration than the segmented periodogram and has many other useful properties such as interpretability of the time and frequency marginals, instantaneous frequency and group delay, and  weak finite support \citep[cf.][]{Sandsten2016}.
These properties however come at a cost in the form of non-disappearing cross-terms or interference terms \citep[e.g.][]{Sandsten2016}. To understand the nature of these terms, consider a signal with two components $X$ and $Y$. The quadratic superposition principle implies
that the pre-periodogram of the signal $X+Y$ is given by
\[
J_T^{(X+Y)}\big(\tfrac{t}{T},\lambda\big)
=J_T^{(X)}\big(\tfrac{t}{T},\lambda\big)
+J_T^{(Y)}\big(\tfrac{t}{T},\lambda\big)
+2\mathop{\mathrm{Re}} J_T^{(X,Y)}\big(\tfrac{t}{T},\lambda\big).
\]
Here the first two terms on the right-hand side are the pre-periodograms of the two signal components while the last term is the so-called cross-term or interference term with $J_T^{(X,Y)}\big(\tfrac{t}{T},\lambda\big)$ being the Fourier transform of the covariance estimates $X(t-k/2)\,Y(t+k/2)$, $k\in\znum$. These interference terms are largest halfway between the two signal components and can dominate the local behavior of the pre-periodogram. Due to their oscillating nature they can be smoothed out to some extent by kernel smoothing with sufficiently large bandwidths. To our knowledge, our method is the first that is designed to actually control for cross-terms allowing to exploit the useful properties of the pre-periodogram in practice. This is further explained in section 3.

The pre-periodogram $J_T(u,\lambda)$ is asymptotically unbiased but for consistency it requires smoothing in both time and frequency direction. For this, let $\Kt$ and $\Kf$ be two smoothing kernels and let $\btT$ and $\bfT$ be two bandwidths for the time and the frequency direction, respectively. Then the kernel smoothing estimator of the time-varying spectral density at rescaled time $u\in[0,1]$ and frequency $\lambda\in[-\pi,\pi]$ is given by
\begin{equation}
\label{eq:fw}
\hatfT(u,\lambda)
=\SSS{\frac{1}{C}}\sum_{s,j}\Kf\Big(\SSS{\frac{\lambda-\lam_j}{\bfT}}\Big)\,
\Kt\Big(\SSS{\frac{u-s/T}{\btT}}\Big)\,\JT\big(\tfrac{s}{T},\lambda_j\big),
\end{equation}
where $\lambda_j=\frac{\pi j}{T}$ for $j=1-T,\ldots,T$ denote the Fourier frequencies and $C=\lsum_{s,j}\Kf\big((\lambda-\lam_j)/\bfT\big)\,\Kt\big((u-s/T)/\btT\big)$ is the normalization constant. The properties of \eqref{eq:fw} have been investigated in the setting of empirical spectral processes in \citet{Dahlhaus2009a} and \citet{Dahlhaus2009b}. In particular, it has been shown that, under suitable conditions on the rates of the smoothing bandwidths, the estimator is asymptotically normal \citep[Theorem 3.2, Example 4.1 of][]{Dahlhaus2009a}. 
In the following, we refer to \eqref{eq:fw} as the non-adaptive estimator.

%% file: Sec3a-propag.tex
\section{Propagation--separation approach in the time--frequency plane}

A well-known problem in the application of kernel estimators is the selection of appropriate bandwidths as the quality of the resulting estimates depends critically on the chosen bandwidth \citep[e.g.][]{Eichler2011}. For time-varying spectral estimators, various asymptotic results are available \citep[see e.g.][]{Dahlhaus2009a, Dahlhaus2009b}, but determining the corresponding optimal smoothing bandwidths in practice is still an open problem. As an alternative to standard bandwidth selection methods, we propose an iterative algorithm to determine at each point in the time--frequency plane the shape of the smoothing kernel data-adaptively. 
Our method is based on the propagation--separation approach by \citet{Polzehl2006}. Starting with small bandwidths for the initial estimates, the bandwidths are increased in each iteration to allow smoothing over larger regions of homogeneity (propagation) while detection of differences between estimates of the preceding iteration leads to penalization and reduction of kernel weights to stop further smoothing (separation).

\begin{figure}[tb]
\begin{center}
  \includegraphics[width=\linewidth]{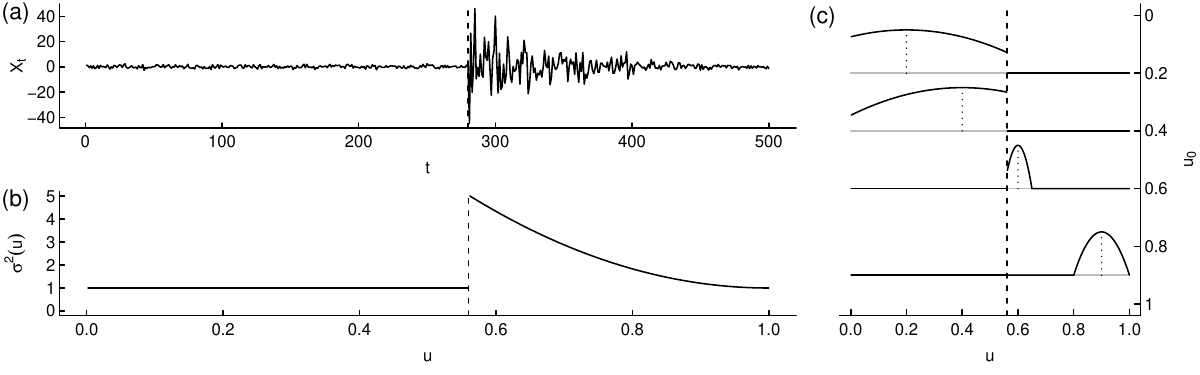}
  \caption{\small{(a) Time series $X_t$ with structural break at $t_{\text{break}}=280$ (dashed lines) and
  (b) time-varying variance $\sigma^2\big(\tfrac{t}{T}\big)$;
  (c) optimal smoothing kernels at locations $u_0\in\{0.2,0.4,0.6,0.9\}$.}}
\label{fig:example}
\end{center}
\end{figure}

To illustrate the main idea, consider the white noise signal $X_t$, $t=1,\ldots,500$, in Figure \ref{fig:example}\,(a) with time-varying variance (Fig.~\ref{fig:example}\,(b)) and a structural break at time $t_{\text{break}}=280$. As the variance is constant for time points $1\leq t\leq t_{\text{break}}$, the smoothing bandwidth in time direction can be chosen as large as possible, but due to the structural break smoothing should not extend across time $t_{\text{break}}$. The corresponding (non-normalized) smoothing kernels in time direction about time points $u_0=0.2$ and $u_0=0.4$ (in rescaled time) are depicted as the upper two curves in Figure \ref{fig:example}\,(c). For times $t_{\text{break}}<t\leq 500$, the variance changes smoothly over time and the bandwidths of the smoothing kernels needs to be adapted accordingly; the two lower curves in Figure \ref{fig:example}\,(c) show the smoothing kernels about the time points $u_0=0.6$ and $u_0=0.9$. Finally we note that since the spectral density is constant over frequencies the smoothing kernels should be unconstrained with maximal bandwidth in frequency direction.

\subsection{The algorithm} \label{algorithm}

For data-adaptive kernel estimation of the time-varying spectral density, we consider a sequence of
weighted averages
\begin{equation}
\label{eq:estimatorfinal}
\hatfkT(u_r,\lambda_i)
=\tfrac{1}{\hatNkT(r,i)}\lsum_{s,j}\WkT_{r,i}(s,j)\,\JT\big(u_s,\lambda_j\big),\quad k=0,\ldots,k_\text{max},
\end{equation}
where $u_r=\tfrac{r}{T}$ and $\lambda_i=\tfrac{2\pi i}{T}$ for $r,i=1,\ldots,T$, $\JT$ is the pre--periodogram given in \eqref{eq:ppgram} and $\hatNkT(r,i)=\sum_{s,j}\WkT_{r,i}(s,j)$ is the sum of weights. The weights $\WkT_{r,i}(s,j)$ determine the shape of the local smoothing kernel at the point $(u_r,\lambda_i)$ and are computed iteratively by the algorithm described below.

The key step in the construction of the data-adaptive kernel weights is a penalization for deviations from local homogeneity. More precisely, we measure differences between the spectral estimates at points $(u_r,\lambda_i)$ and $(u_s,\lambda_j)$ in the time-frequency plane by the discrepancy
\begin{equation}
\label{eq:statisticpen}
\Delta\big(\hatfkkT{k}(u_r,\lambda)i),\hatfkkT{k}(u_s,\lambda_j)\big)
= \SSS{\frac{\hatNkT(r,i)}{4\pi\hatMkT(r,i)}}
\bigg(\SSS{\frac{\hatfkT(u_r,\lambda_i)-\hatfkT(u_s,\lambda_j)}{\barfkT(u_r,\lambda_i)}}\bigg)^2,
\end{equation}
where $\hatMkT_{r,i}=\sum_{s,j}\WkT_{r,i}(s,j)^2$ and $\barfkT(u_r,\lambda_i)$ in the denominator is a local average of the $\hatfkT$ about the point $(u_r,\lambda_i)$. Under the assumption of homogeneity, the above discrepancy is asymptotically $\chi^2$ distributed with one degree of freedom and thus provides a measure of the violation of homogeneity independent of the underlying unknown spectral density. For a brief discussion of the effect of penalization on the asymptotic distribution of the adaptive estimator we refer to the supplement \citep{vDE18supp}. Discrepancies larger than the $90\%$ quantile of the $\chi^2_1$ distribution are judged as violation of local homogeneity while for smaller values of the discrepancy a penalty factor $P^{(k)}_{r,i}(s,j)$ is obtained by application of a decreasing kernel $K_P$.

One major problem in the application of the penalization are cross terms that distort the time-frequency signal in the form of highly oscillating and possible negative components. While the effect of such cross terms is diminished by smoothing over large enough local regions, it has a serious impact on penalization and can lead to isolation of points in the time-frequency plane. To prevent such deterioration over iterations of the algorithm we introduce a relaxation towards a locally smoothed version of the current estimate, where the amount of shift is determined by the local signal-to-noise ratio being a measure for the stability of the current estimate.

For locally stationary processes with smoothly changing (that is, not piecewise constant) spectral densities, another problem arises from the usual bias at peaks and troughs \citep[e.g.][]{Dahlhaus1990}. To prevent accumulation of bias over iterations, we also add a relaxation towards the most recent estimate. This relaxation step is controlled by the integrated penalty factor, which indicates saturation in the growth of the effective smoothing region.

Below we first present the complete algorithm in detail. In the next subsection, we then provide further explanations and additional comments.

\begin{list}{}{\labelwidth1em\leftmargin1em\labelsep0em\topsep0.25em plus 0.5ex%
\itemsep0.25em plus 0.5ex\parsep0em}
\item[{\bfseries Initialization ($\boldsymbol{k=0}$).\ }]\mbox{}\\
For initial bandwidths $\btT^{(0)}$ and $\bfT^{(0)}$ compute initial kernel estimates
\[
\hat f^{(0)}(u_r,\lambda_i)
=\tfrac{1}{\#N^{(0)}(r,i)}\sum_{(s,j) \in  B^{(0)}(r,i)}\Kf\big(\tfrac{\lam_j-\lam_i}{\bfT^{(0)}}\big)\,\Kt\big(\tfrac{u_s-u_r}{\btT^{(0)}}\big)\,\JT(u_s,\lam_j)
\]
and local averages
\[
\bar f^{(0)}(u_r,\lambda_i)
=\tfrac{1}{\# B^{(0)}(r,i)}\sum_{(s,j)\in B^{(0)}(r,i)}\hat f^{(0)}(u_s,\lambda_j),
\]
where $N^{(0)}(r,i)=\lsum_{s,j}\Kf((\lam_j-\lam_i)/\bfT^{(0)})\,\Kt((u_s-u_r)/\btT^{(0)})$ and $B^{(k)}(r,i) =\{(s,j):\,|u_s-u_r|<\btT^{(k)},|\lambda_j-\lambda_i|<\bfT^{(k)}\}$ for $k=0,\ldots,k_{\text{max}}$.\\
{\em Parameters: } smoothing kernels $\Kt$ and $\Kf$ (default$\Kt(x)=\Kf(x)=6\big(\tfrac{1}{4}-x^2\big)\,1_{[-1/2,1/2]}(x)$)
\item[{\bfseries Iteration $\boldsymbol{k-1\to k}$.\ }]\mbox{}
\begin{alphlist}
\item
Increase kernel bandwidths: $\btT^{(k)}=\alpha_{\text{t}}\,\btT^{(k-1)}$ and $\bfT^{(k)}=\alpha_{\text{f}}\,\bfT^{(k-1)}$.\\
{\em Parameters:} growth rates $\alphat$ and $\alphaf$ (default values $\alphat=\alphaf=1.2$)
\savecounter{alphcount}
\end{alphlist}
\item[{\bfseries (1) Penalty step.}]\mbox{}
\begin{alphlist}
\restorecounter{alphcount}
\item
From discrepancies $\Delta\big(\hatfkkT{k-1}(u_r,\lambda_i),\hatfkkT{k-1}(u_s,\lambda_j)\big)$ (defined in \eqref{eq:statisticpen}) between previous estimates $\hatfkkT{k-1}(u_r,\lambda_i)$ and $\hatfkkT{k-1}(u_s,\lambda_j)$ compute the penalties for the deviation from homogeneity
\[
\tilde P^{(k)}_{r,i}(s,j)
=K_P\Big(\big(\hatfkkT{k-1}(u_r,\lambda_i),
\hatfkkT{k-1}(u_s,\lambda_j)\big)\Big),
\]
where $K_P$ is a decreasing non-negative penalty kernel with bounded support $[0,c_P]$;\\
constrain final penalty $P^{(k)}_{r,i}(s,j)$ to be radially non-increasing:
\[
P^{(k)}_{r,i}(s,j)
=\min\big(\tilde P^{(k)}_{r,i}(s,j),(1-\alpha)\,P^{(k)}_{r,i}(s-1,j)+\alpha\,P^{(k)}_{r,i}(s-1,j-1)\big)
\]
for points $(s,j)$ such that $(s,j)=\beta\,(s-1,j-\alpha)$ for some $\beta>1$ and $\alpha\in[0,1]$ and
\[
P^{(k)}_{r,i}(s,j)
=\min\big(\tilde P^{(k)}_{r,i}(s,j),(1-\alpha)\,P^{(k)}_{r,i}(s-1,j)+\alpha\,P^{(k)}_{r,i}(s-1,j-1)\big)
\]
for points $(s,j)$ such that $(s,j)=\beta\,(s-\alpha,j-1)$ for some $\beta>1$ and $\alpha\in[0,1]$.\\
{\em Parameters:} cut-off $c_P$ for penalties (default $c_P=\chi^2_{1,0.9}$);
penalty kernel $K_P$ (default $K_P(x)=1-(x/c_P)^2$ for $x\leq c_P$ and zero elsewhere)
\item
Compute intermediate estimator
\begin{equation}
\label{eq:estimatorad}
\tildefkT(u_r,\lambda_i)
=\tfrac{1}{\tildeNkT(r,i)}\lsum_{(s,j) \in \in B^{(k)}(r,i)}\tildeWkT_{r,i}(s,j)\,\JT\big(u_s,\lambda_j\big)
\end{equation}
from adapted kernel weights
\begin{equation}
\label{eq:weightsad}
\tildeWkT_{r,i}(s,j)=\Kf\big(\tfrac{\lambda_i-\lambda_j}{\baf^{(k)}}\big)\,
\Kt\big(\tfrac{u_s-u_r}{\bt^{(k)}}\big)\,P^{(k)}_{r,i}(s,j),
\end{equation}
where $\tildeNkT(r,i)=\sum_{s,j}\tildeWkT_{r,i}(s,j)$ and $\tildeMkT(r,i)=\sum_{s,j} \tildeWkT_{r,i}(s,j)^2$.
\item
Compute total measure of penalization
\[
\pi^{(k)}(r,i)=\tfrac{1}{2}\,\big(\max(\pi^{(k)}_{\text{t}+}(r,i),\pi^{(k)}_{\text{t}-}(r,i))+
\max(\pi^{(k)}_{\text{f}+}(r,i),\pi^{(k)}_{\text{f}-}(r,i))
\]
from directional penalization measures in time directions
\begin{align*}
\pi^{(k)}_{\text{t}+}(r,i)&=\tfrac{1}{C_{\text{t}+}}\,\lsum_{(s,j)\in B^{(k)}(r,i)}P^{(k)}_{r,i}(s,j)\,\tfrac{(s-r)^+}{\sqrt{(s-r)^2+(j-i)^2}},\\
\pi^{(k)}_{\text{t}-}(r,i)&=\tfrac{1}{C_{\text{t}-}}\,\lsum_{(s,j)\in B^{(k)}(r,i)}P^{(k)}_{r,i}(s,j)\,\tfrac{(s-r)^-}{\sqrt{(s-r)^2+(j-i)^2}}
\end{align*}
with $C_{\text{t}\pm}=\lsum_{(s,j)\in B^{(k)}(r,i)}\tfrac{(s-r)^\pm}{\sqrt{(s-r)^2+(j-i)^2}}$ 
and similarly $\pi^{(k)}_{\text{f}+}(r,i)$ and $\pi^{(k)}_{\text{f}-}(r,i)$ in frequency directions. Additionally compute the overall penalization at $(u_r,\lambda_i)$ as
\[
\pi^{(k)}_{\text{all}}(u_r,\lambda_i)
=\tfrac{1}{\#B^{(k)}(r,i)}\,\lsum_{(s,j)\in B^{(k)}(r,i)} P^{(k)}_{r,i}(s,j).
\]
\item
compute the signal-to-noise ratio
\begin{equation}
s^{(k)}(u_r,\lam_i)=\frac{\bar f^{(k)}_{\text{hom}}(u_r,\lambda_i)}{\tilde\sigma^{(k)}_{\text{hom}}(u_r,\lambda_i)}
\label{snrat}
\end{equation}
from local mean and variance of $\tilde f^{(k)}$
\begin{align*}
\bar f^{(k)}_{\text{hom}}(u_r,\lambda_i)
&=\tfrac{1}{\# B_{\text{hom}}(r,i)}\lsum_{(s,j)\in B_{\text{hom}}(r,i)}\tilde f^{(k)}(u_s,\lam_j),\\
\tilde\sigma^{(k)}_{\text{hom}}(u_r,\lambda_i)^2
&=\tfrac{1}{\# B_{\text{hom}}(r,i)}\lsum_{(s,j)\in B_{\text{hom}}(r,i)}
\big(\tilde f^{(k)}(u_s,\lam_j)-\bar f^{(k)}_{\text{hom}}(u_r,\lambda_i)\big)^2
\end{align*}
over local region of homogeneity
\[
B_{\text{hom}}(r,i)
=\{(s,j):\,|u_s-u_r|<\btT^{(0)},|\lambda_j-\lambda_i|<\bfT^{(0)},P^{(k)}_{r,i}(s,j)>\tfrac{1}{2}\}.
\]
\savecounter{alphcount}
\end{alphlist}
\item[{\bfseries (2) Relaxation I:\ }] cancellation of negative cross-terms
\begin{alphlist}
\restorecounter{alphcount}
\item
If $\tildefkT(u_r,\lambda_i)<\tildefkkT{k-1}(u_r,\lambda_i)<0$ set
\[
\tildefkT(u_r,\lambda_i)=\tildefkkT{k-1}(u_r,\lambda_i);
\]
similarly set $\tildeNkT(r,i)$, $\tildeMkT(r,i)$, $\bar f^{(k)}_{\text{hom}}(u_r,\lambda_i)$, and $s^{(k)}(u_r,\lam_i)$ to their previous values $\tildeNkkT{k-1}(r,i)$, $\tildeMkkT{k-1}(r,i)$, $\bar f^{(k-1)}_{\text{hom}}(u_r,\lambda_i)$, and $s^{(k-1)}(u_r,\lam_i)$, respectively.
\savecounter{alphcount}
\end{alphlist}
\item[{\bfseries (3) Relaxation II:\ }] shift towards smoothed version of intermediate estimates
\begin{alphlist}
\restorecounter{alphcount}
\item
Compute local averages of $\tilde f^{(k)}(u,\lambda)$, the kernel normalization, and the total penalization measure
\begin{align}
\bar f^{(k)}_{\text{ave}}(u_r,\lambda_i)
&=\tfrac{1}{\#N^{(0)}(r,i)}\lsum_{(s,j)\in B^{(0)}(r,i)} \Kf\big(\tfrac{\lam_j-\lam_i}{\bfT^{(0)}}\big)\,\Kt\big(\tfrac{u_s-u_r}{\btT^{(0)}}\big)\tilde f^{(k)}(u_s,\lambda_j),
\label{ftildeav}\\
\bar N^{(k)}_{\text{ave}}(u_r,\lambda_i)
&=\tfrac{1}{\#N^{(0)}(r,i)}\lsum_{(s,j)\in B^{(0)}(r,i)}\Kf\big(\tfrac{\lam_j-\lam_i}{\bfT^{(0)}}\big)\,\Kt\big(\tfrac{u_s-u_r}{\btT^{(0)}}\big)\tilde N^{(k)}(s,j),
\label{Ntildeav}\\
\bar\pi^{(k)}_{\text{ave}}(r,i)
&=\tfrac{1}{\# N^{(0)}(r,i)}\lsum_{(s,j)\in B^{(0)}(r,i)}\Kf\big(\tfrac{\lam_j-\lam_i}{\bfT^{(0)}}\big)\,\Kt\big(\tfrac{u_s-u_r}{\btT^{(0)}}\big)\pi^{(k)}(s,j).
\label{nutildeav}
\end{align}
\item
From the signal-to-noise ratio $s^{(k)}(u_r,\lambda_i)$ defined in (e) compute the relaxation parameter 
\[
\theta^{(k)}(r,i)=\min\big(c_s/s^{(k)}(u_r,\lambda_i),1\big),
\]
for all points $(u_r,\lambda_i)$ for which the stability condition $\tilde N^{(k)}(r,i)\geq \alpha_{\text{f}}\,\alpha_{\text{t}}\,\bar N^{(k)}_{\text{ave}}(r,i)$ holds and set $\theta^{(k)}(r,i)=1$ if the stability condition is violated; In that case, we additionally set $\tilde N^{(k)}(r,i)=N^{(k-1)}(r,i)$ and $\tilde M^{(k)}(r,i)=M^{(k-1)}(r,i)$.\\
{\em Parameters:} threshold $c_s$ for the signal-to-noise ratio (default $c_s=2$)
\item
Obtain new estimate $\tildefrelkT$ as weighted average of $\tildefkT$ and smoothed version $\bar f^{(k)}_{\text{ave}}$
\[
\tildefrelkT(u_r,\lambda_i)=\big(1-\theta^{(k)}(r,i)\big)\,\tilde f^{(k)}(u_r,\lambda_i)
+\theta^{(k)}(r,i)\,\bar f^{(k)}_{\text{ave}}(u_r,\lambda_i);
\]
similarly shift the local mean and the total penalization measure towards their smoothed versions
\begin{align*}
\barfrelkT(u_r,\lambda_i)&=\big(1-\theta^{(k)}(r,i)\big)\,\bar f^{(k)}_{\text{hom}}(u_r,\lambda_i)
+\theta^{(k)}(r,i)\,\bar f^{(k)}_{\text{ave}}(u_r,\lambda_i),\\
\pi^{(k)}_{\text{rel}}(r,i)&=\big(1-\theta^{(k)}(r,i)\big)\,\pi^{(k)}(r,i)
+\theta^{(k)}(r,i) \bar\pi^{(k)}_{\text{ave}}(r,i).
\end{align*}
\savecounter{alphcount}
\end{alphlist}
\item[{\bfseries (4) Relaxation III:\ }] shift towards estimate from previous iteration
\begin{alphlist}
\restorecounter{alphcount}
\item
Compute final estimate $\hatfkT$ as weighted average of $\tildefrelkT$ and $\hatfkkT{k-1}$
\[
\hatfkT(u,\lambda)=(1-K_p\big(\pi^{(k)}_{\text{rel}}(r,i) \,c_P\big))\,\tildefrelkT(u_r,\lambda_i)
+K_p\big(\pi^{(k)}_{\text{rel}}(r,i)\,c_P\big)\,\hatfkkT{(k-1)}(u_r,\lambda_i),
\]
similarly $\bar f^{(k)}(u_r,\lambda_i)$, $\NkT(r,i)$, and $\MkT(r,i)$ are obtained from
$\barfrelkT(u_r,\lambda_i)$, $\tildeNkT(r,i)$, and $\tildeMkT(r,i)$, respectively, and their versions from the previous iteration.
\end{alphlist}
\item[{\bfseries Loop.\ }]
Repeat step (1) to (4) until the average overall penalty over all design points,
\begin{equation}\label{eq:stop}
\bar\pi^{(k)}_{\text{all}}=\tfrac{1}{T^2}\lsum_{r,i=1}^{T} \pi^{(k)}_{\text{all}}(u_r,\lambda_i),
\end{equation}
where $\pi^{(k)}_{\text{all}}(u_r,\lambda_i)$ is the overall penalty defined in (d), indicates that the majority of all smoothing kernels are shrinking, that is, until
\[
\bar\pi^{(k)}_{\text{all}}<\frac{1}{\alpha_{\text{f}}\alpha_{\text{t}}},
\]
or until the maximal number $k_\text{max}$ of iterations with $\bfT^{(k_\text{max})} =\btT^{(k_\text{max})}\approx 1$ is reached. In case the algorithm terminates by \eqref{eq:stop}, estimates are returned for
\[
k_{\text{final}}=\max\big\{k\big|\bar\pi_{\text{all}}^{(k)}\geq \big(\tfrac{1}{4}+\tfrac{3}{4}\,\alphat\,\alphaf\big)^{-1}\big\}.
\]
\end{list}

%

%% file: Sec3b-details.tex
\subsection{Further details}

We now provide further details on the initialization parameters and on the various steps. 

\subsubsection{Parameters for initialization}

\begin{list}{}{\labelwidth0em\leftmargin0em\labelsep0em\topsep0.25em plus 0.5ex%
\itemsep0.25em plus 0.5ex\parsep0em}
\item[{\em Initial bandwidth parameters $\btT^{(0)}$ and $\bfT^{(0)}$:\ }] 
the choice of initial bandwidth faces the usual trade-off of being able to retrieve details in the signal while not having too strong distortions due to noise. The algorithm offers default values set to $\bfT^{(0)}=\btT^{(0)}= \sqrt{\log{T}^{1.9}/ 2\pi T}$, which are conservative compared to the CLT condition underlying the distribution of the penalty statistic \citep[Assumption 1.1(ii). of][]{vDE18supp}. Additionally the algorithm has the option to automatically improve the initial bandwidths if increasing them slightly could reduce the percentage of negative initial estimates.
\item[{\em Choice for kernel functions $K_t$, $K_f$ and $\Kst$:\ }] the default smoothing kernels $K_t$ and $K_f$ of the algorithm have been shown to yield the smallest mean squared error \citep{Dahlhaus1996b} in case of local homogeneity. In combination we have found the concave penalty kernel $\Kst$ most appropriate. The cut-off value for the penalty kernel is based on the asymptotic distribution of the penalty statistic under local homogeneity. 
\end{list}

\subsubsection{Penalty step}

\begin{list}{}{\labelwidth0em\leftmargin0em\labelsep0em\topsep0.25em plus 0.5ex%
\itemsep0.25em plus 0.5ex\parsep0em}
\item[{\em Step (b) and (e):\ }]
for the relative squared error in the discrepancy, we use a local average of the estimates over a small region to improve stability of the penalization step. In order to prevent bias from high curvature or structural breaks the local average is taken only over points that are judged as belonging to the same homogeneous region by the discrepancy.
\item[{\em Step (d):\ }]
the penalization measures in step (d) give an indication about the growth of the effective smoothing region, which is required for controlling relaxation towards previous estimates as well as for stopping of the algorithm. In order to allow effective smoothing regions to grow further even in the close neighborhood of structural breaks that lead to strong penalization on one side, penalization is measured along the time and frequency directions separately. For this, penalties are weighted by the component of their normalized vectors in the direction of interest.
\end{list}

\subsubsection{Relaxation I and II}

The first two relaxation steps control for the effect of cross-terms and reduce their presence iteratively.
Cross-terms often lead to highly oscillating positive and negative spikes in the pre-periodogram that make recovering of the spectral signal very difficult. Empirically, areas that are dominated by cross-terms can be identified by low signal-to-noise ratio. Relaxation II stabilizes the areas affected by cross-terms by applying a local smoothing based on the local signal-to-noise ratio.
\begin{list}{}{\labelwidth0em\leftmargin0em\labelsep0em\topsep0.25em plus 0.5ex%
\itemsep0.25em plus 0.5ex\parsep0em}
\item[{\em Step (f):\ }]
negative cross-terms can lead to negative estimates that over iterations can destabilize the estimation. We therefore limit the effect by replacing negative estimates by their least observed versions prior to Relaxation II.
\item[{\em Step (g):\ }]
we use a signal-to-noise ratio to assess the reliability of the estimates. The signal-to-noise ratio is obtained by the ratio of a local average and a estimate of the local standard deviation. The local region over which these measures are computed is chosen such that it only includes points within a small neighborhood, as given by the starting bandwidths, that have been identified by the discrepancy measure to potentially belong to the same homogeneous region 
\item[{\em Steps (h) to (j):\ }]
full weight is provided to a local average if the signal to noise ratio is less than $c_s$ or a large cross-term is detected. Large cross-terms can be identified as points of which the sum of weights is extremely low compared to the local average sum of weights of the same step. By imposing local smoothing in these regions we allow to distinguish signal from noise iteratively enabling the detection of breaks as well as smooth patterns.
\end{list}
\subsubsection{Relaxation III and early stopping}
\begin{list}{}{\labelwidth0em\leftmargin0em\labelsep0em\topsep0.25em plus 0.5ex%
\itemsep0.25em plus 0.5ex\parsep0em}
\item[{\em Step (k):\ }]
the purpose of the last relaxation step is to stabilize the estimate if the total penalization measure indicates that no further smoothing over larger regions is required. In each step, the bandwidths of the neighborhood over which can be smoothed are increased. Meaning that, unless the smoothing region covers a homogeneous area, the penalty kernel becomes less pronounced as the distance kernels flatten. This causes a bias that we control for by bending  the estimator to its previous estimator according to the total penalization measure. 
\item[{\em Early stopping:\ }]
in principle, the algorithm continues until the entire plane is searched for each point, that is, until $\bfT^{(k_\text{max})} =\btT^{(k_\text{max})}\approx 1$.  The algorithm stops earlier if \eqref{eq:stop} is satisfied, i.e., if smoothing kernels are shrinking for the majority of design points.
Further smoothing will then no longer lead to improved estimates and the returned estimates are specified by the largest $k$ for which $\bar\pi^{(k)}_{\text{all}}$ is bounded away from the stopping threshold $1/\alphat\alphaf$.
\end{list}

%% file: Sec5-examp.tex
\section{Simulations}\label{Examples}

In this section, we first illustrate the our data-adaptive estimation method by application to simulated data from three processes that cover three types of possible scenarios: structural break with otherwise constant spectrum, smooth time-varying spectrum, and time-varying spectrum that also exhibits a structural break. In all examples, the default parameters are taken to demonstrate that our method can be expected to work well for a wide range of processes without requiring process-specific choice of parameters.
Furthermore, to investigate the performance of our data-adaptive spectral estimator quantitatively, we compare it in a simulation study to a kernel spectral estimator with global bandwidth chosen such that the mean square error is minimized. As the optimal choice of the bandwidth depends on the unknown spectral density, this estimator can be seen as an oracle estimator. Throughout this section, we therefore refer to this estimator as the oracle estimator.

\subsection{Structural break white noise}\label{examplestructbreak}

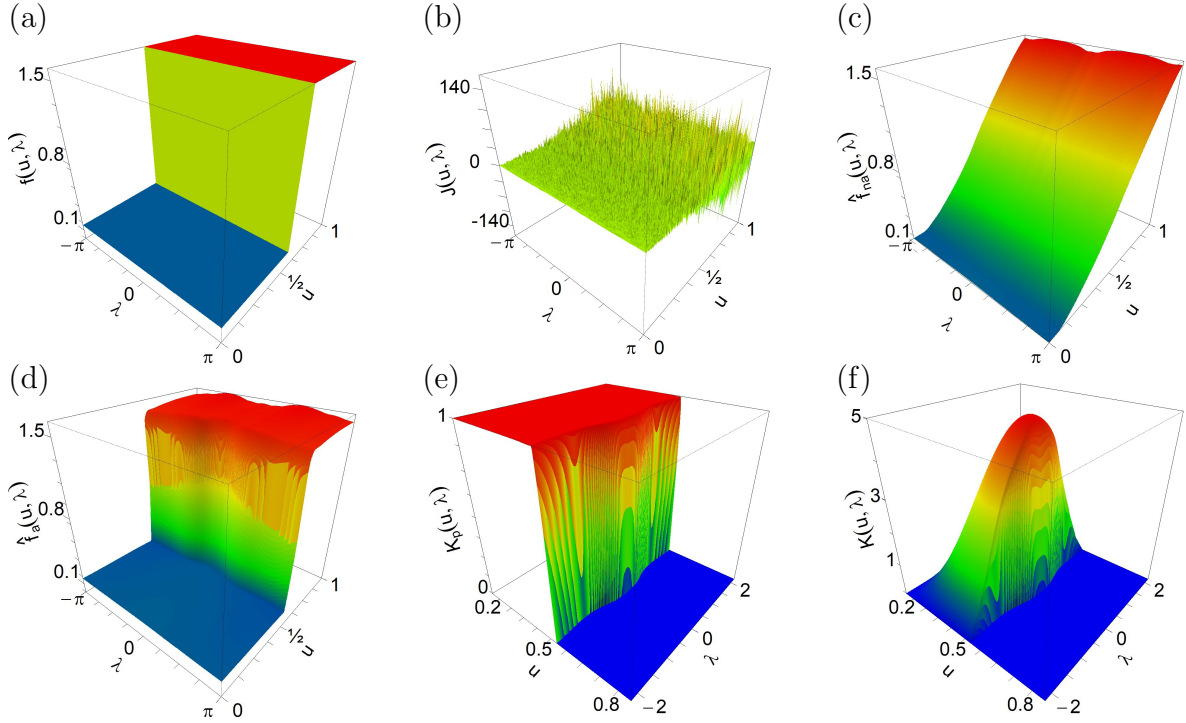
\begin{figure}[tb]
\begin{center}
\setlength{\unitlength}{1truecm}
\begin{pspicture}(0,0)(16.5,10.2)
  \put( 0.0,4.7){\includegraphics[width=4.8cm]{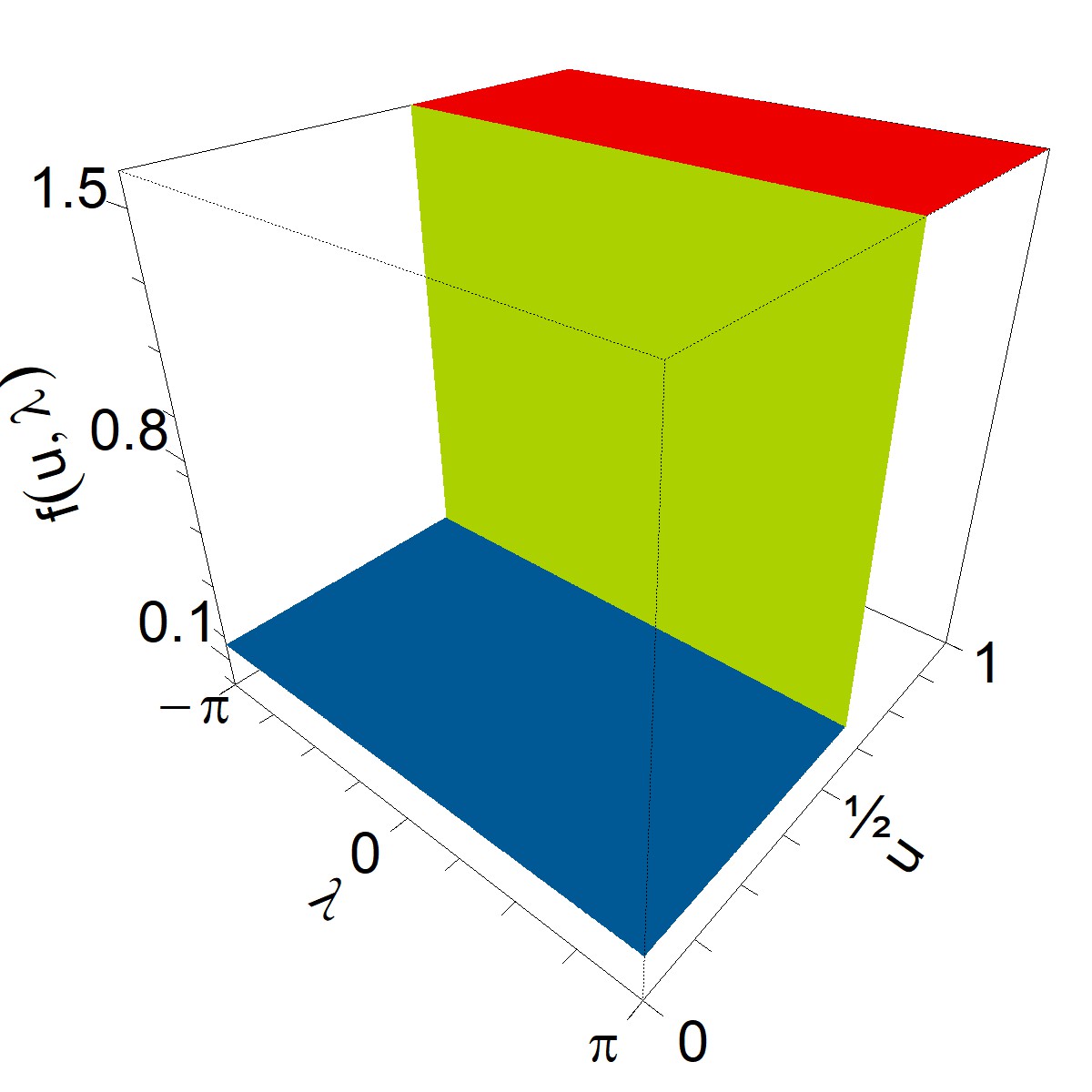}}
  \put( 0.0,9.3){(a)}
  \put( 5.5,4.7){\includegraphics[width=4.8cm]{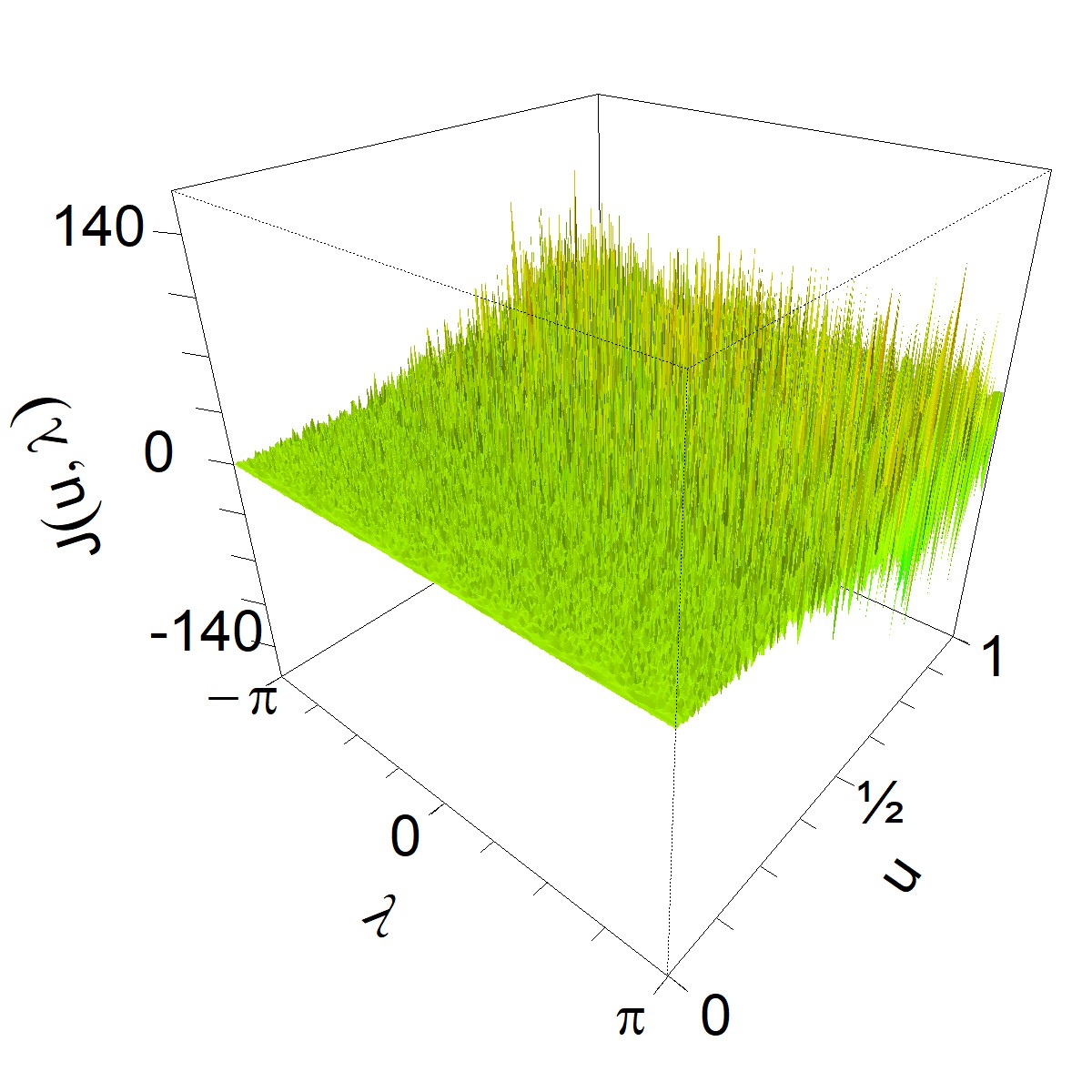}}
  \put( 5.5,9.3){(b)}
  \put(11.0,4.7){\includegraphics[width=4.8cm]{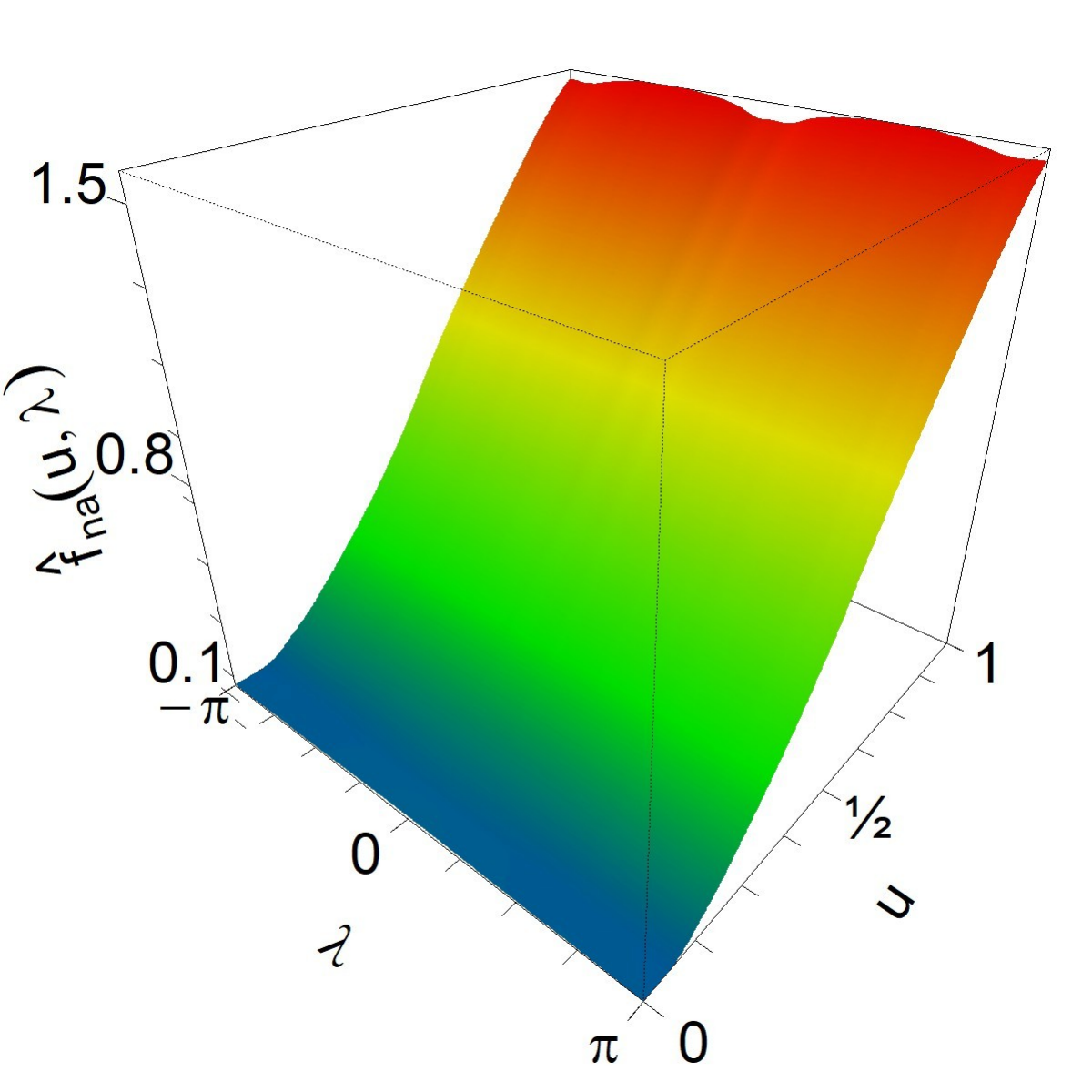}}
  \put(11.0,9.3){(c)}
  \put( 0.0,0.0){\includegraphics[width=4.8cm]{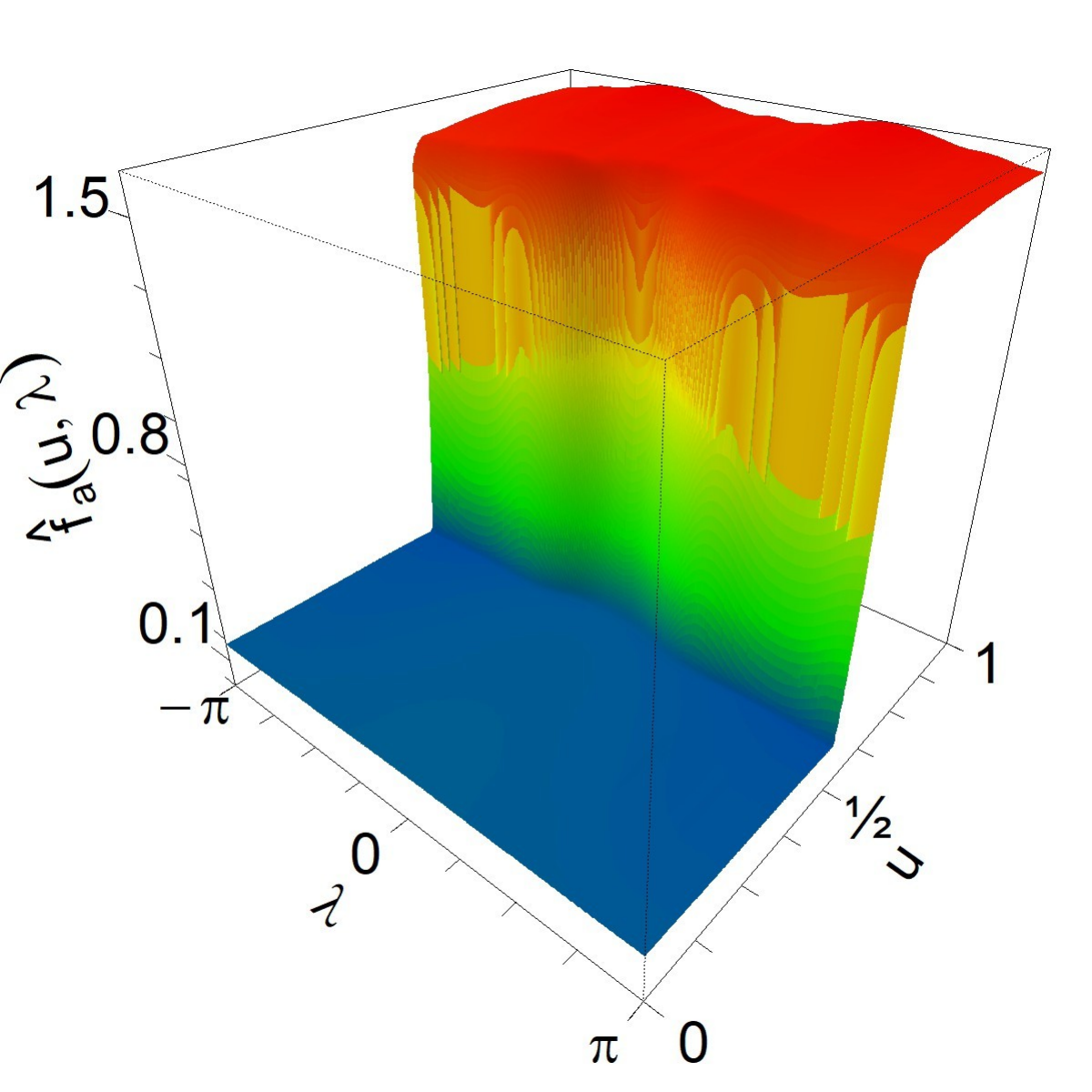}}
  \put( 0.0,4.5){(d)}
  \put( 5.5,0.0){\includegraphics[width=4.8cm]{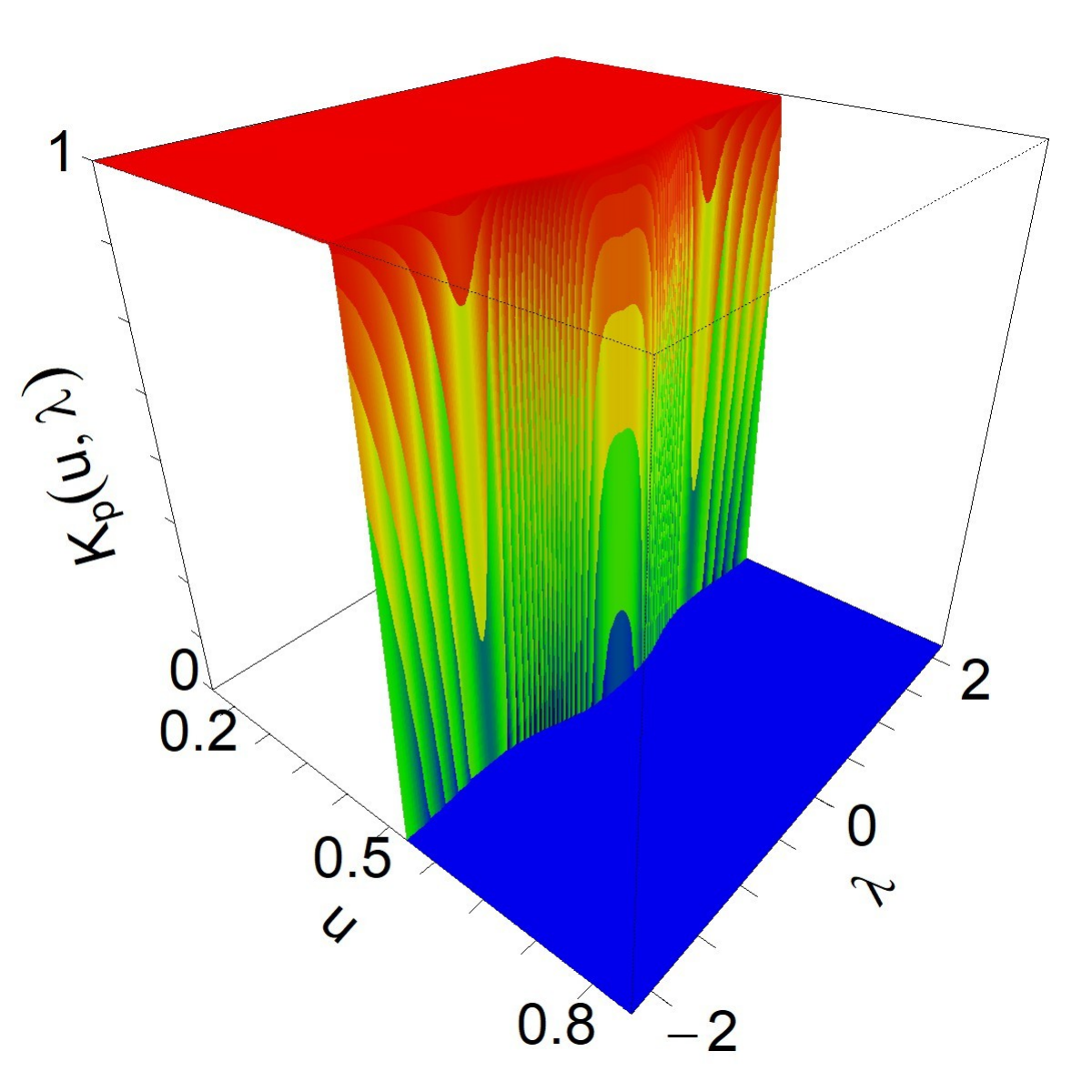}}
  \put( 5.5,4.5){(e)}
  \put(11.0,0.0){\includegraphics[width=4.8cm]{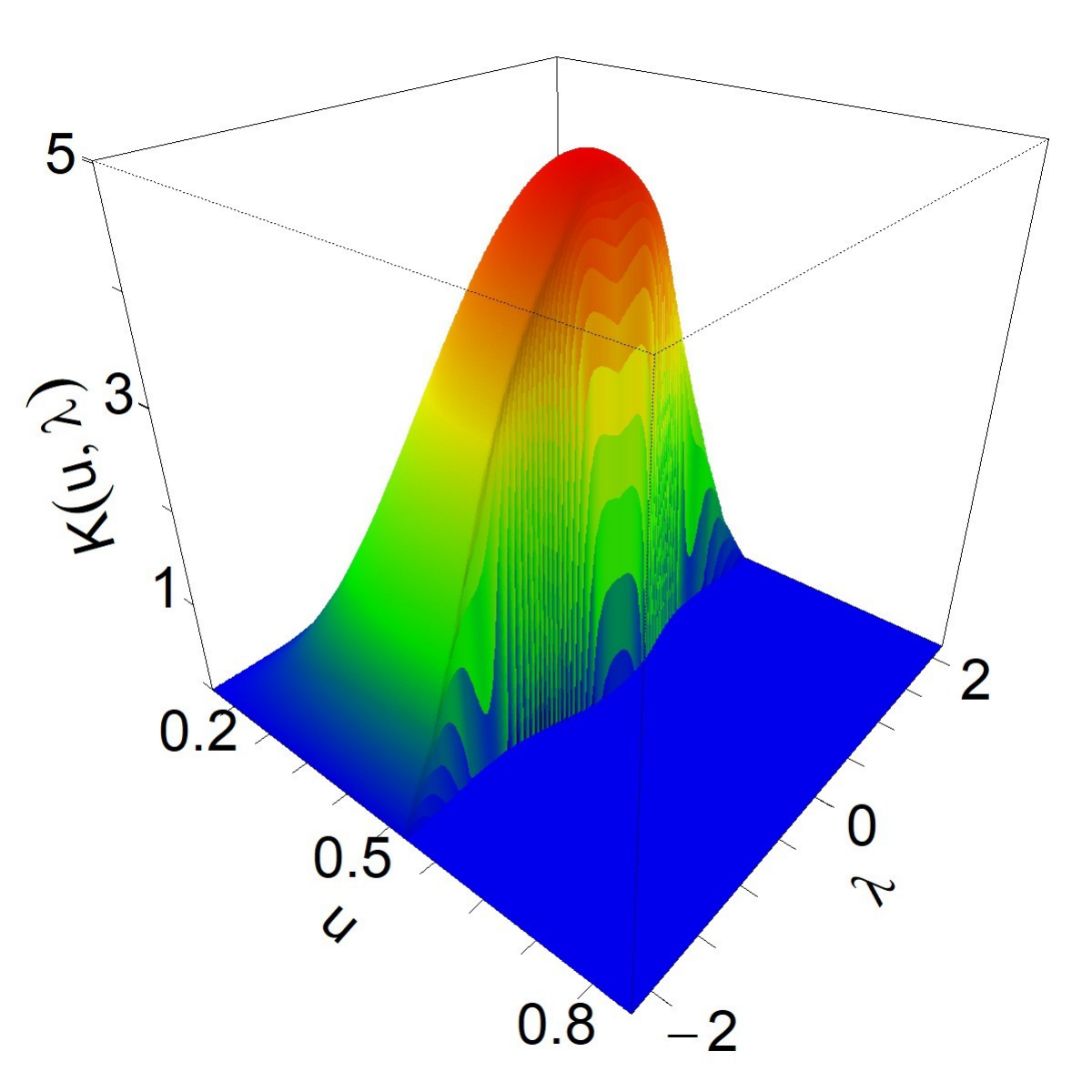}}
  \put(11.0,4.5){(f)}
\end{pspicture}
\caption{White noise with structural break: (a) true spectrum, (b) pre--periodogram, (c) nonadaptive estimate, (d) adaptive estimate, (e) penalty kernel and (f) smoothing kernel (scaled by a factor $10^5$) of the adaptive estimate at $(u,\lambda)=\big(\tfrac{1}{2},0\big)$.}
\label{fig:wn1}
\end{center}
\end{figure}

The first process that we consider is a white noise process with an upward shift in the variance of the process at some time $t_0$. More specifically, data $X_1,\ldots,X_T$ were simulated from the following model
\[
X_{t}=\begin{cases}
\veps_t&\text{ if }t\leq t_0\\
\sqrt{10}\,\veps_t&\text{ if }t>t_0
\end{cases}
\]
with $\veps_t\iid\normal(0,1)$ and $t_0=0.5625\,T$ where $T=1024$. The corresponding time-varying spectral density is given by
\begin{equation}
f_X(u, \lambda) = \SSS{\frac{1}{2 \pi}}\,1_{\{u \leq t_0/T\}}
+\SSS{\frac{10}{2 \pi}}\,1_{\{u >t_0/T\}}
\end{equation}
and is depicted in Figure \ref{fig:wn1}\,(a). Figure \ref{fig:wn1}\,(b) shows the corresponding pre-periodogram. Compared with the ordinary periodogram, it exhibits much more variation which completely blurs the piecewise constant form of the density. The algorithm stopped after $k_{\text{final}}=k_{\text{max}}=14$ iterations, that is, the search region was extended until covering the entire plane. The resulting data-adaptive estimate is given in Figure \ref{fig:wn1}\,(d). It clearly shows two levels for the spectral density. Figures \ref{fig:wn1}\,(e) and (f) depict, respectively, the final penalty and the adaptive kernel for the point $(u,\lambda)=\big(\tfrac{1}{2},0\big)$ in the middle of the plane.  Once the break is detected the penalty kernel forces the weights down to zero. This results in an asymmetric smoothing kernel that is `cut off' and thus succeeds in separating the areas on both sides of the break (compare with the example in Figure \eqref{fig:example}). To illustrate the effect of penalization, Figure \ref{fig:wn1}\,(c) shows a non-adaptive kernel estimate as given in \eqref{eq:fw} computed with the same bandwidths as used in the final step of our iterative method. Not surprisingly, the presence of the break is completely smoothed out.

\begin{figure}[tb]
\begin{center}
\includegraphics[width=12cm]{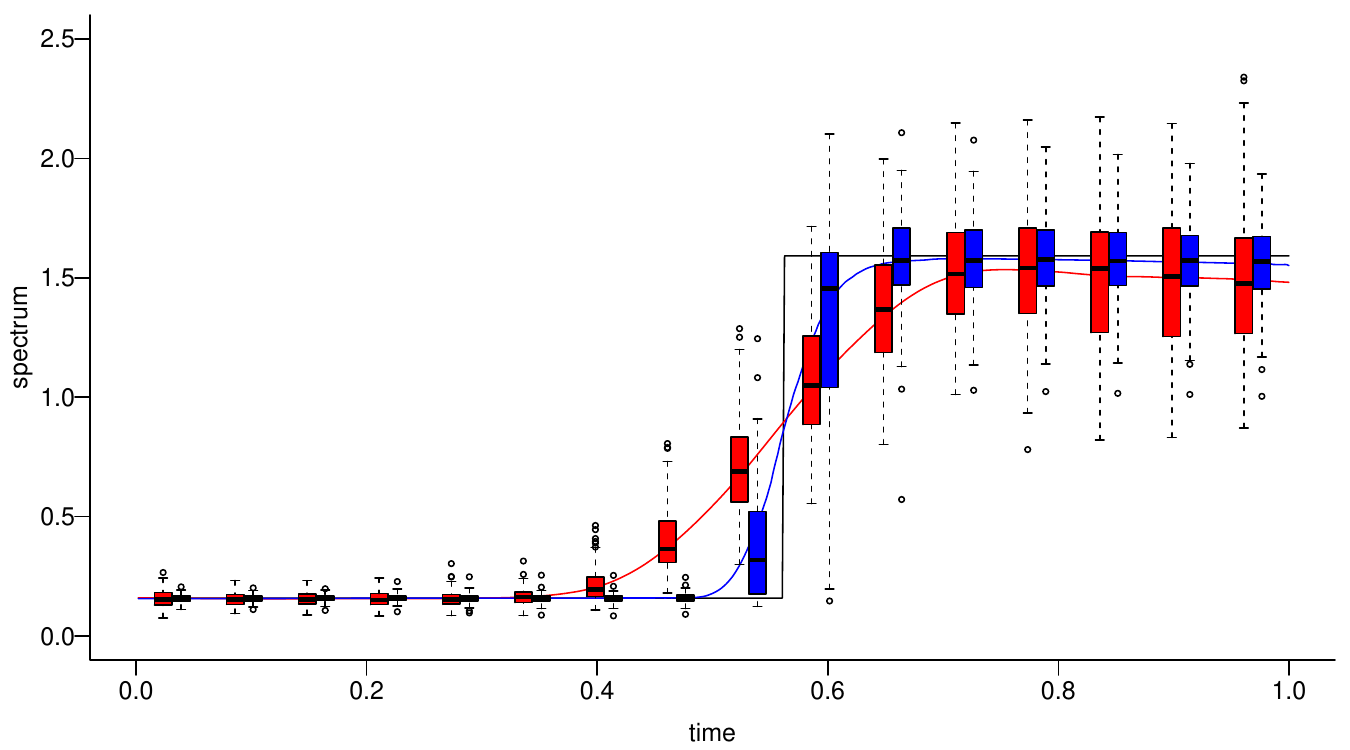}
\caption{Comparison of spectral estimates for process $X_t$ (white noise with structural break) over 100 repetitions: mean curves and boxplots for the adaptive estimator $\fmax$ (blue) and the oracle kernel estimator $\fnaopt$ (red) over time $u$ for frequency $\lambda=0$; the true spectral density is added in black.}
\label{fig:wn2}
\end{center}
\end{figure}
In order to examine the performance of our estimator, we compared it with the oracle estimator over 100 repetitions of process $X_t$. Figure $\ref{fig:wn2}$ depicts the mean curves and pointwise boxplots at frequency $\lambda=0$ for different time points $u$. 
It can be observed that close the break the adaptive estimator varies much more than the oracle estimator. This can be explained by the fact that the adaptive estimator clearly detects the break but detect its location slightly before or slightly after the true break point in the different repetitions. The oracle estimator does not detect it but simply smooths it out. Away from the break, our estimator is much more precise and is very close to the true spectrum. The good performance of our estimator compared to any non-adaptive kernel estimator with global bandwidths can also be seen from the root integrated mean square errors and the integrated mean absolute errors listed in Table \ref{tab:Errors}.

\begin{table}[tbh]
\vspace*{0cm} 
 \begin{center}
\begin{tabular}{l@{\qquad}rr|rr@{\qquad}rr|rr}
\hline
\hline
&\multicolumn{4}{c}{RMSE}&\multicolumn{4}{c}{MAE}\\
&\multicolumn{2}{c}{$\fmax$}&\multicolumn{2}{c}{$\fnaopt$}&\multicolumn{2}{c}{$\fmax$}&\multicolumn{2}{c}{$\fnaopt$}\\
\hline
Process $X_t$&0.182&(0.044)&0.270& (0.032) &0.092& (0.036) &0.174& (0.027)\\
Process $Y_t$&0.034&(0.008)&0.033& (0.007) &0.021& (0.003) &0.022& (0.004)\\
Process $Z_t$&0.028&(0.004)&0.031& (0.003) &0.020& (0.003) &0.024& (0.003)\\
\hline
\hline
\end{tabular}
\caption{Mean (standard deviation) of the root integrated mean square error (RIMSE) and integrated mean absolute error (IMAE) for the adaptive estimator $\fmax$ and the oracle kernel estimator with optimal bandwidths $\fnaopt$, obtained over 100 repetitions of the three processes $X_t$, $Y_t$, and $Z_t$.}
\label{tab:Errors}
\end{center}
\end{table}

\subsection{Locally stationary series}\label{locstatprocess}
\begin{figure}[tbh]
\begin{center}
\setlength{\unitlength}{1truecm}
\begin{pspicture}(0,0)(16.5,10.2)
  \put( 0.0,4.7){\includegraphics[width=4.8cm]{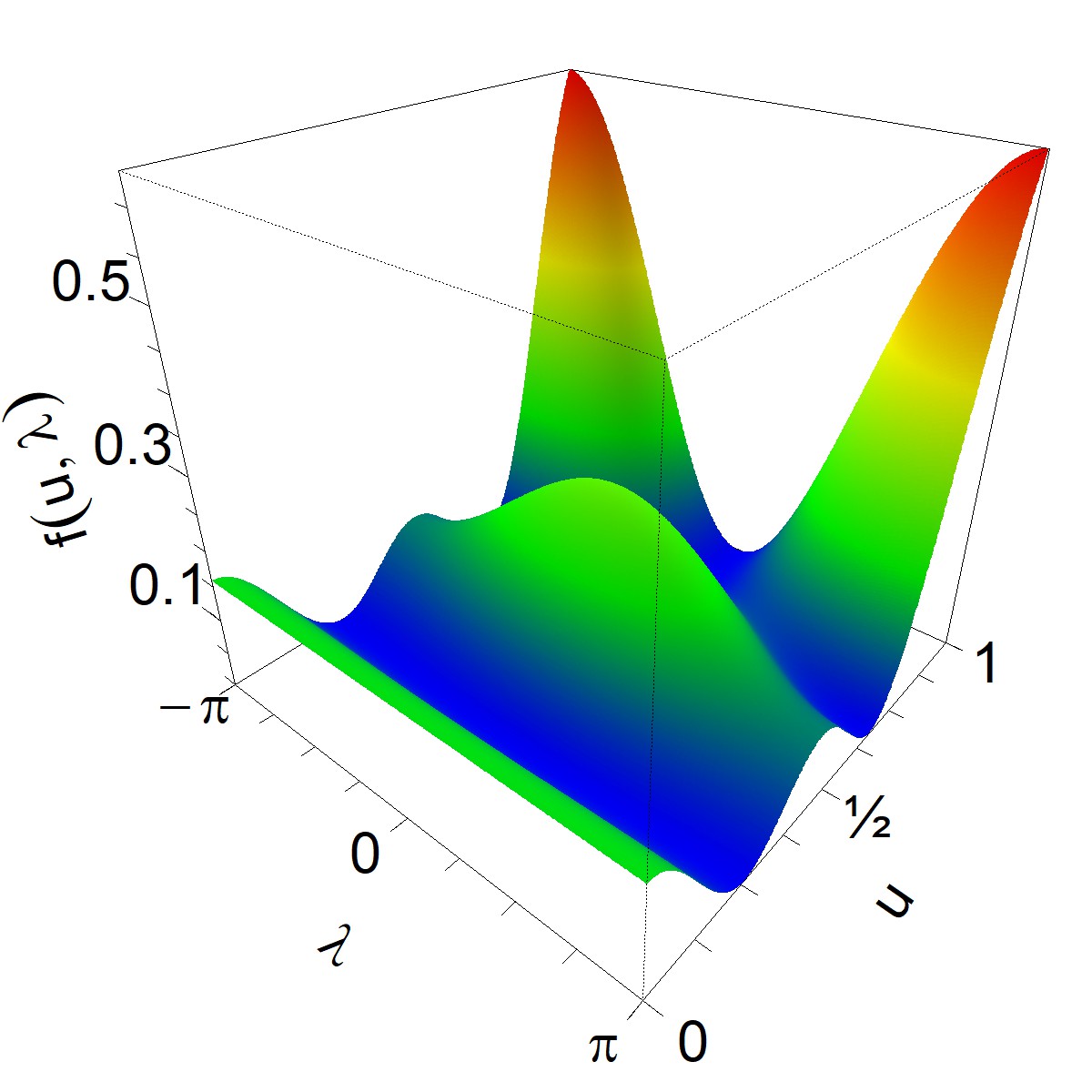}}  
  \put( 0.0,9.3){(a)}                                                       
  \put( 5.5,4.7){\includegraphics[width=4.8cm]{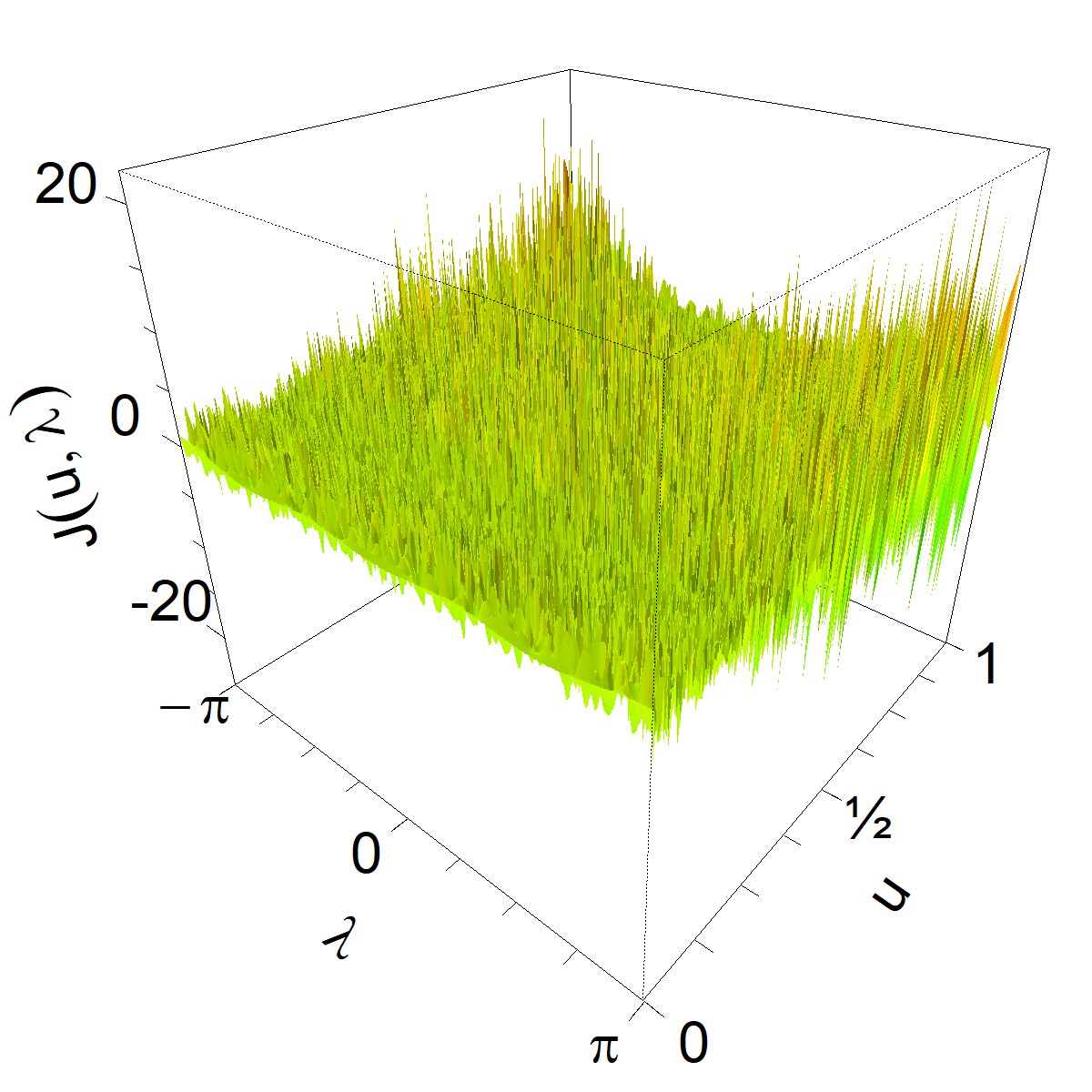}}     
  \put( 5.5,9.3){(b)}                                                       
  \put(11.0,4.7){\includegraphics[width=4.8cm]{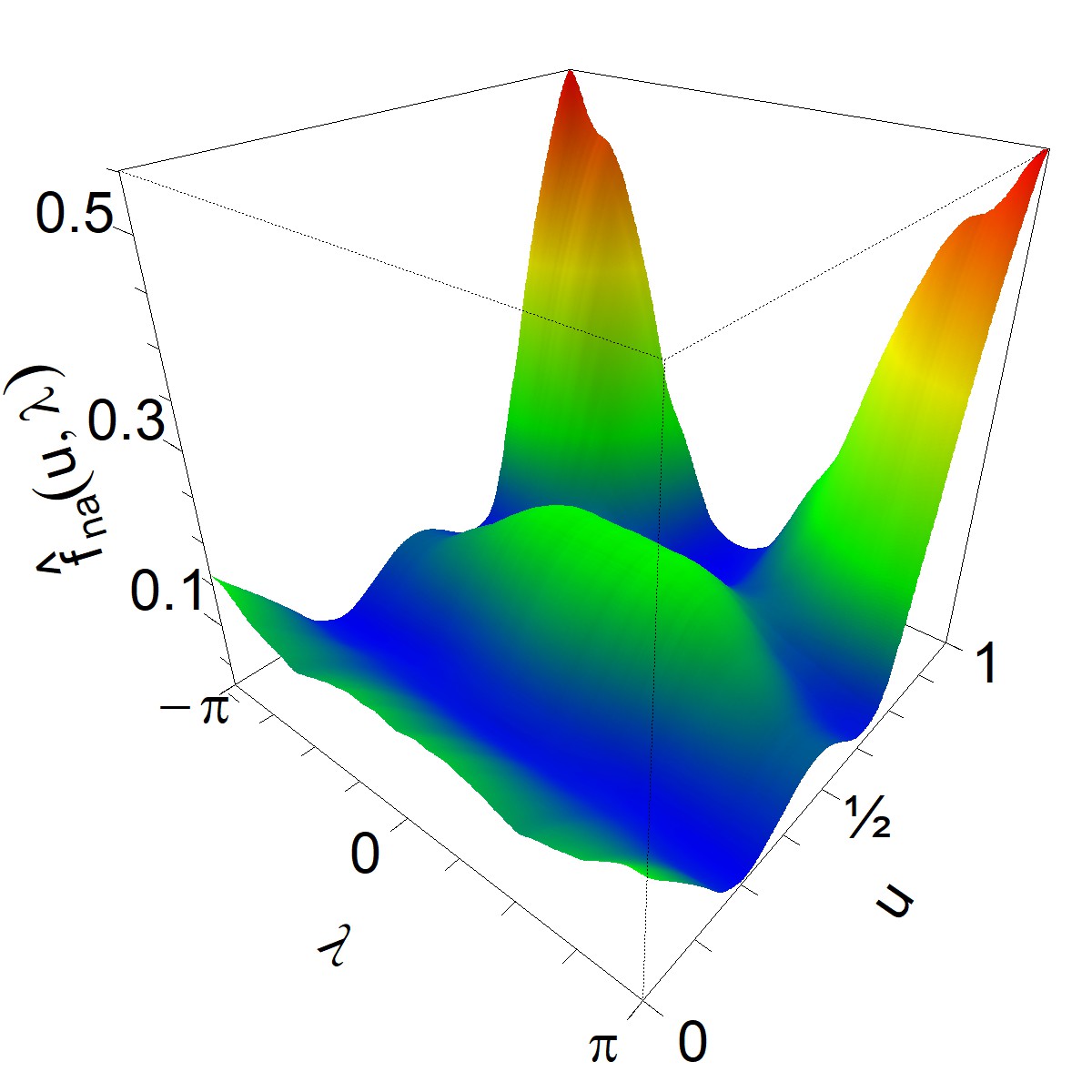}}                  
  \put(11.0,9.3){(c)}                                                       
  \put( 0.0,0.0){\includegraphics[width=4.8cm]{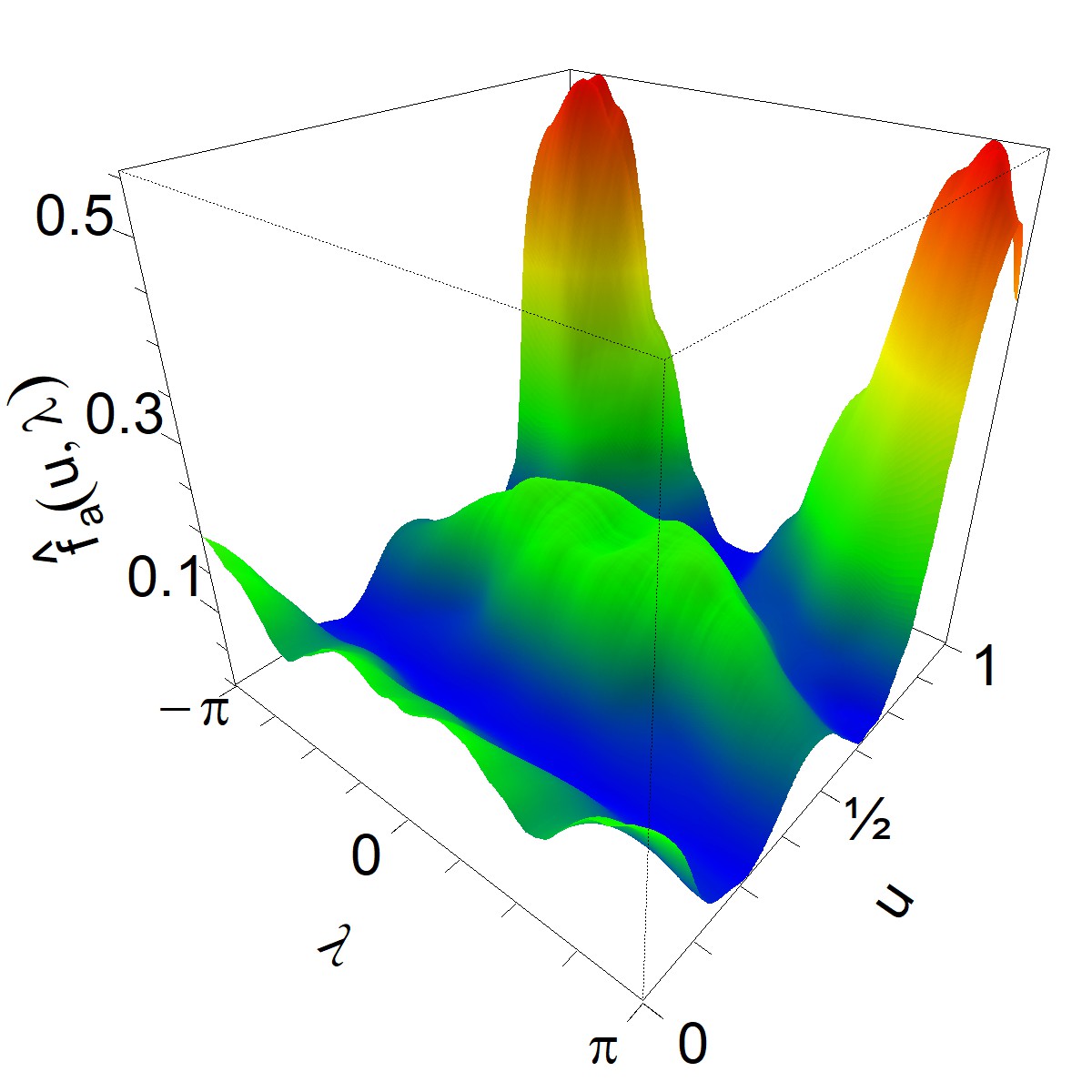}}      
  \put( 0.0,4.5){(d)}                                                       
  \put( 5.5,0.0){\includegraphics[width=4.8cm]{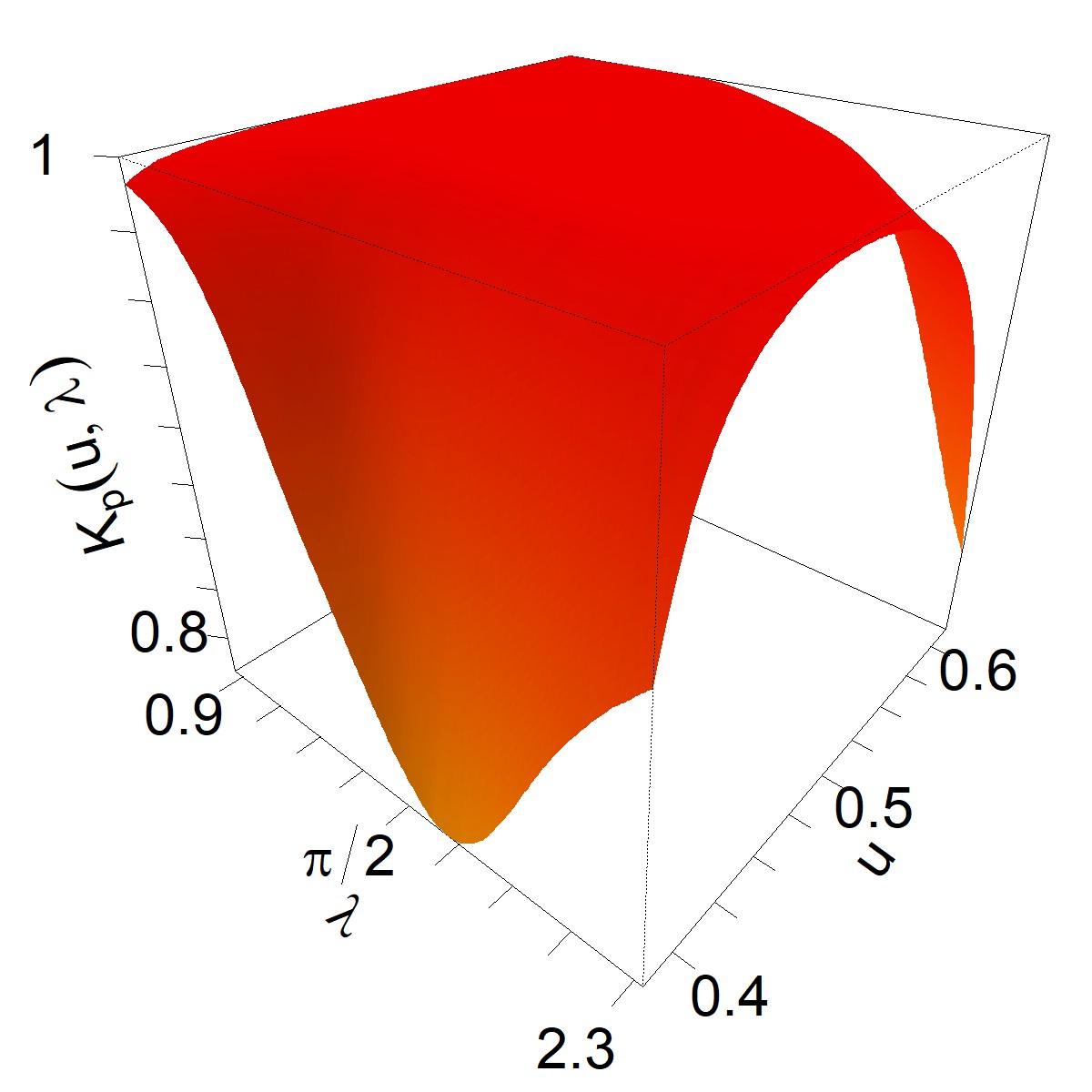}}    
  \put( 5.5,4.5){(e)}                                                       
  \put(11.0,0.0){\includegraphics[width=4.8cm]{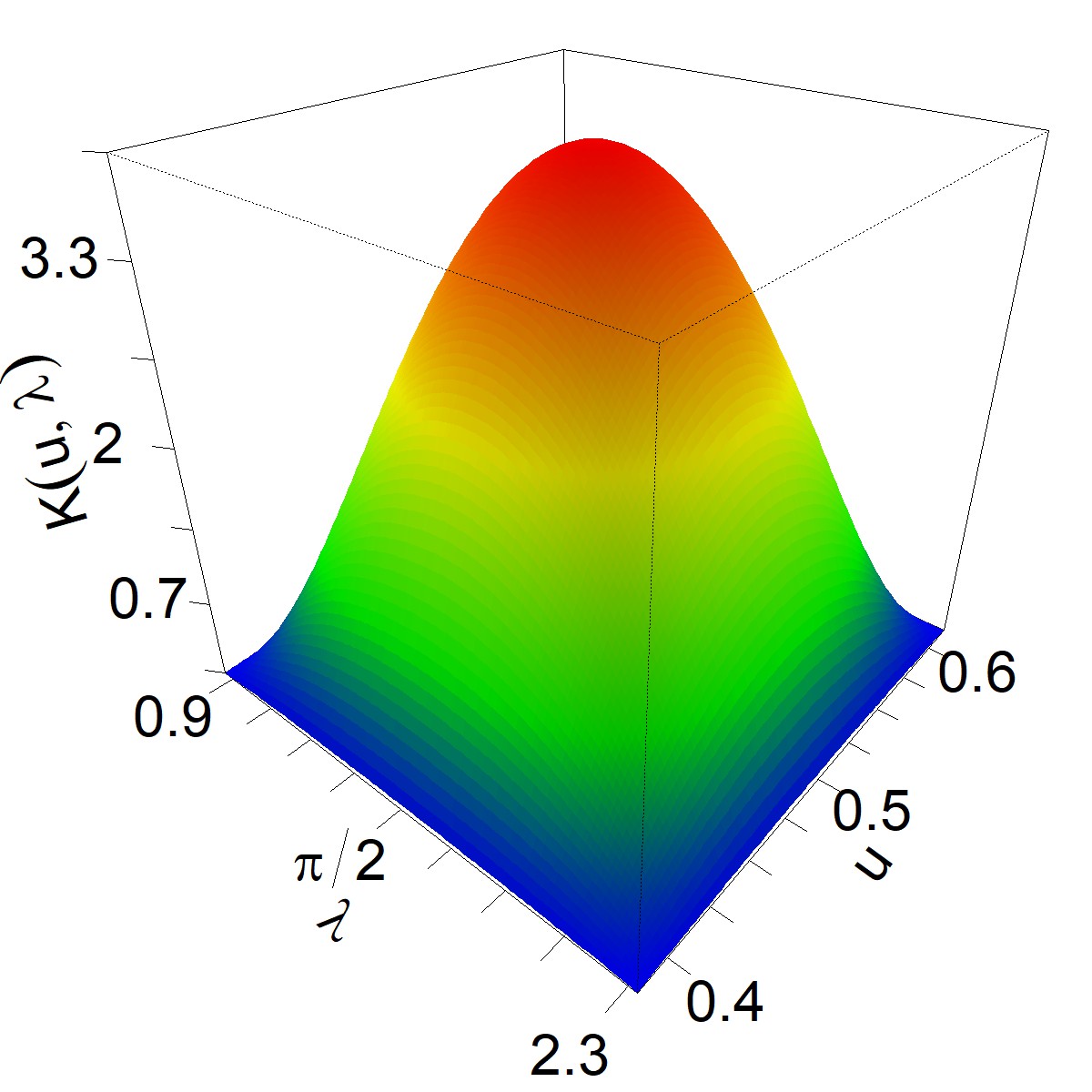}}     
  \put(11.0,4.5){(f)}                                                       
\end{pspicture}
\caption{Locally stationary process: (a) true spectrum, (b) pre--periodogram, (c) nonadaptive estimate, (d) adaptive estimate, (e) penalty kernel and (f) smoothing kernel (scaled by a factor $10^5$) for the adaptive estimate at $(u,\lambda)=\big(\tfrac{1}{2},\tfrac{\pi}{2}\big)$.}
\label{fig:ma1}
\end{center}
\end{figure}
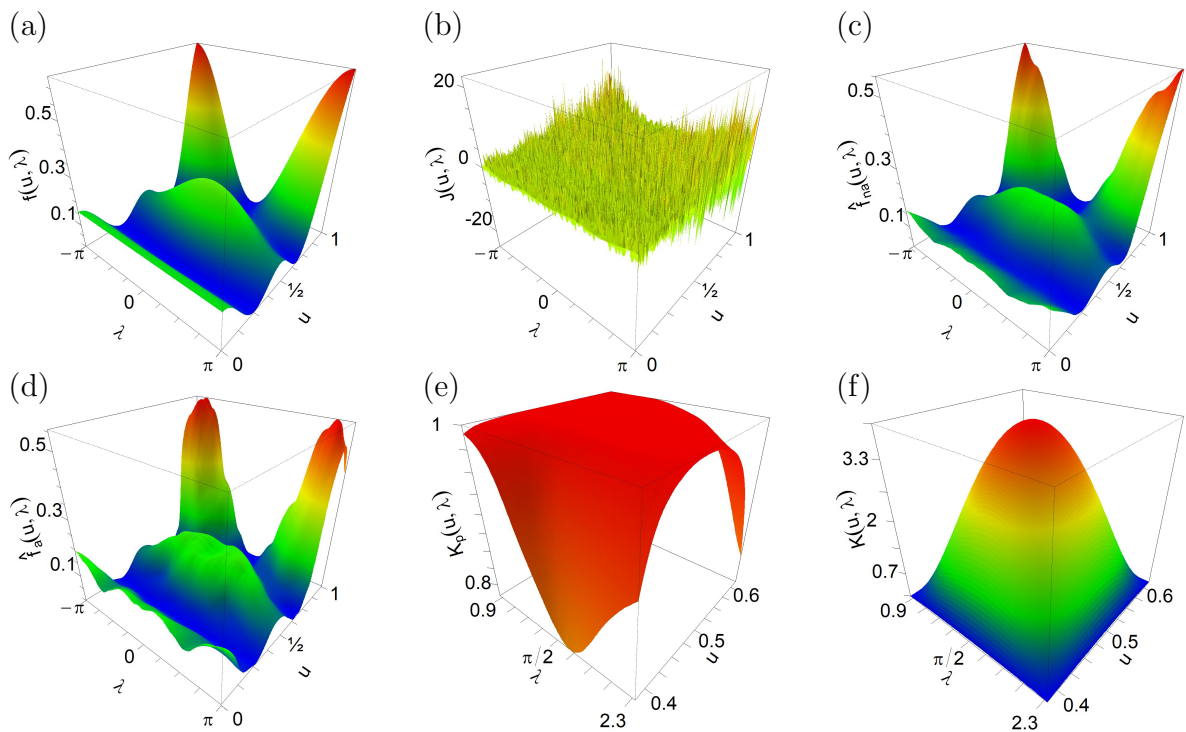

As a second process, we consider a time-varying moving average of order 1 given by
\[
Y_t=\cos\big(2\,\pi\,\tfrac{t}{T}\big)\,\veps_t
-\big(\tfrac{t}{T}\big)^2\,\veps_{t-1}
\]
with $\veps_{t}\iid\normal(0,1)$ and $T=1024$.
The time-varying spectral density of this process, given by
\begin{equation}
\label{ex:locstatdensity}
f(u, \lambda)
=\SSS{\frac{1}{2\pi}} 
\big(\cos(2\pi u)^2-2u^2\cos(2 \pi u)\cos(\lambda)+u^4\big),
\end{equation}
is depicted in Figure \ref{fig:ma1}\,(a). The spectrum exhibits a peak in the middle of the time-frequency plane with smooth hill-sides in frequency direction and steeper ones in time direction. The pre-periodogram is given in Figure \ref{fig:ma1}\,(b) and indicates low signal throughout the plane with notably larger oscillations close to $u=1$.The final estimate $\fmax$ was obtained after $k_{\text{final}}=7$ iterations of the algorithm and is shown in Figure \ref{fig:ma1}\,(d). The estimated spectrum seems to capture the level and curvature of the spectrum very well, even at the time boundaries. In contrast, the non-adaptive kernel estimate $\fna$ with comparable bandwidths  (Fig.~\ref{fig:ma1}\,(c)) gives a smoother impression but seems to out-smooth some of the curvature. The penalty and smoothing kernel of the adaptive estimator at the point $(u,\lambda)=(\tfrac{1}{2},\tfrac{\pi}{2})$ are provided in Figure \ref{fig:ma1}\,(e) and (f). This point is on the smooth hillside of the peak. Barely any penalization is visible which is in line with the smoothness properties of the underlying process. 

\begin{figure}[h!]
\begin{center}
\includegraphics[width=12cm]{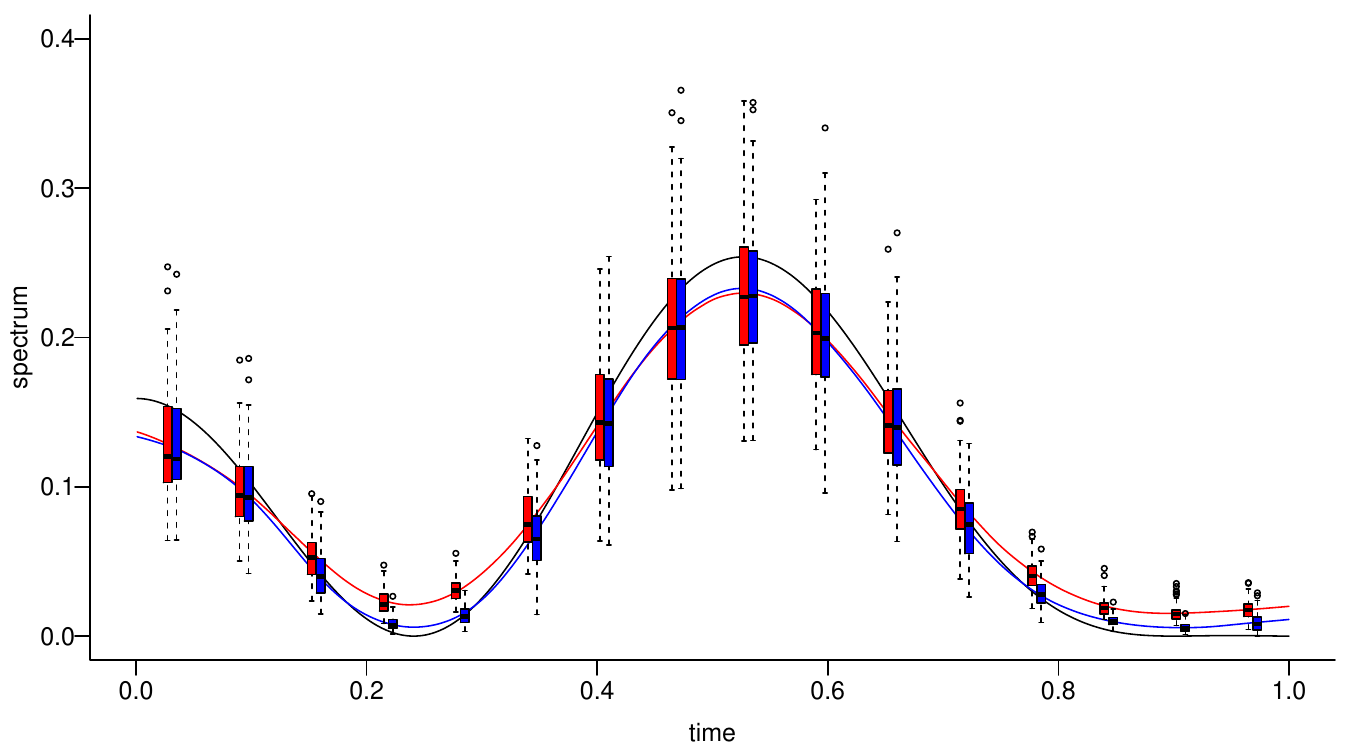}
\caption{Comparison of spectral estimates for the locally stationary process $Y_t$ over 100 repetitions: mean curves and boxplots for the adaptive estimator $\fmax$ (blue) and the oracle kernel estimator $\fnaopt$ (red) over time $u$ for frequency $\lambda=0$; the true spectral density is added in black.}
\label{fig:ma2}
\end{center}
\end{figure}

Figure \ref{fig:ma2} and the second row of Table \ref{tab:Errors} provide the results from the simulation study for process $Y_t$. In this case the comparison with the oracle estimator is very interesting as the true spectral density does not exhibit distinct steep and flat regions but changes in the true spectral density are similar across the entire plane. This suggests that peaks and troughs can be resolved well with one global bandwidth which makes it a very advantageous situation for the oracle estimator.

The error measures in Table \ref{tab:Errors} indicate that both the adaptive and the oracle estimator give comparable results. Figure $\ref{fig:ma2}$ provides more details about the behavior of the two estimators. As can be observed, the oracle estimator shows similar bias at peaks and troughs despite the much smaller variation in the lower regions. In contrast, the bias of the adaptive estimator adjusts to the level of variation and thus is less biased in the troughs. In particular, we note that the true spectral density lies almost always within the range of observed estimates, which is not the case for the oracle estimator. Summarizing we find that although the performance is very similar, the adaptive estimator seems to capture the curvature slightly better than the oracle estimator. We emphasize that the chosen bandwidth of the oracle estimator is computed from the true spectral density and therefore that any estimator with global bandwidths will show larger errors.

\subsection{Structural break in locally stationary series} \label{ma2sbreak}

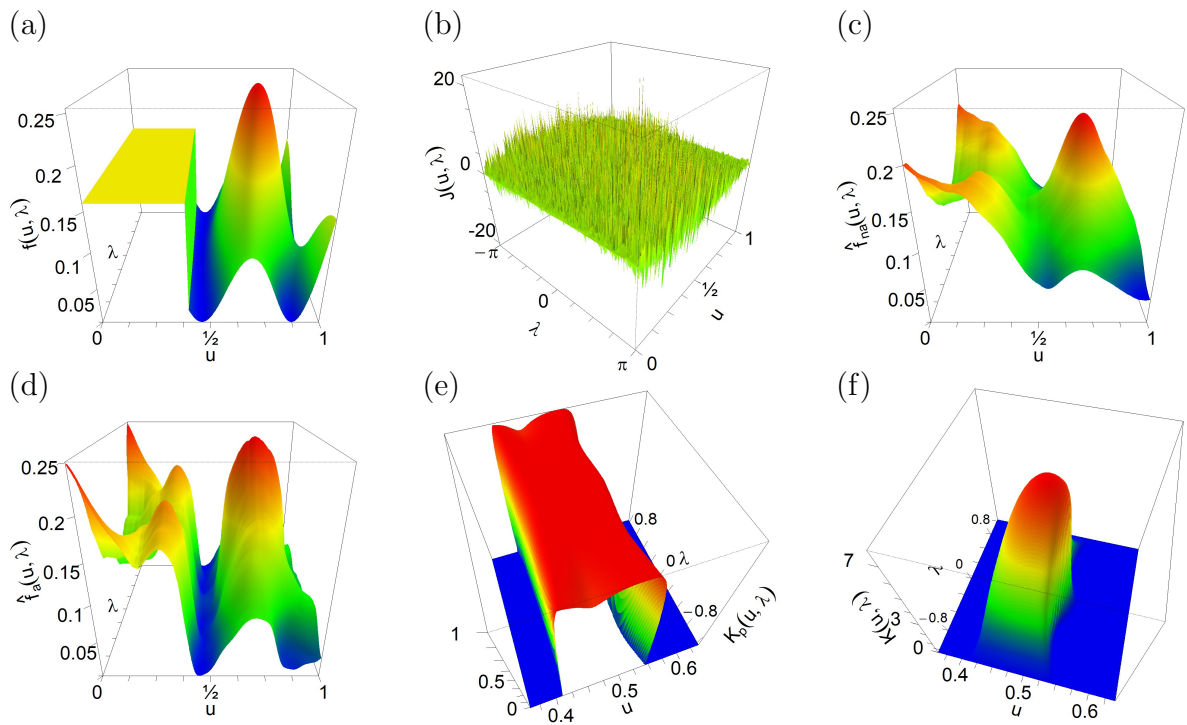
\begin{figure}[tbh]
\begin{center}
\setlength{\unitlength}{1truecm}
\begin{pspicture}(0,0.0)(16.5,10.2)
  \put( 0.0,4.7){\includegraphics[width=4.8cm]{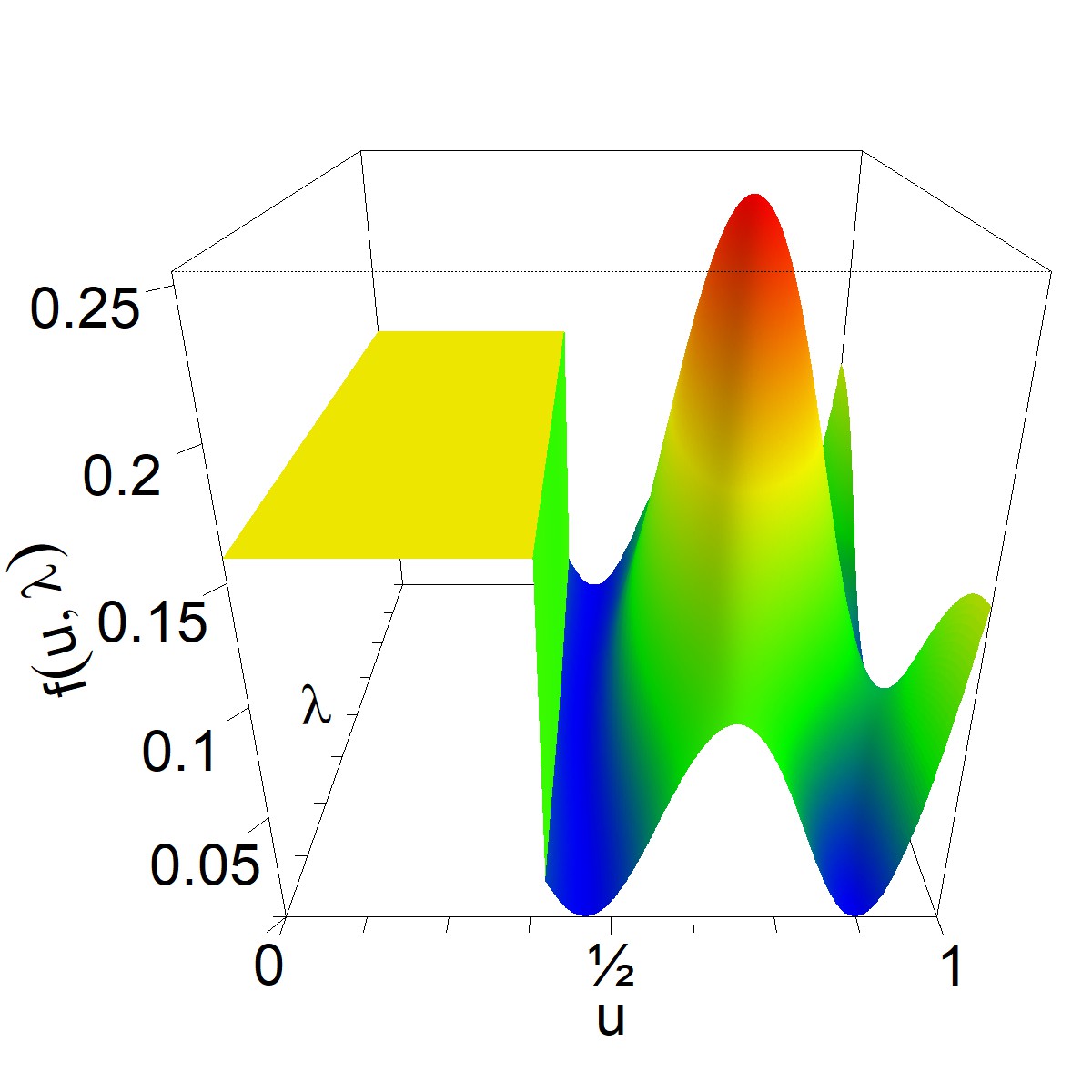}} 
  \put( 0.0,9.3){(a)}                                                       
  \put( 5.5,4.7){\includegraphics[width=4.8cm]{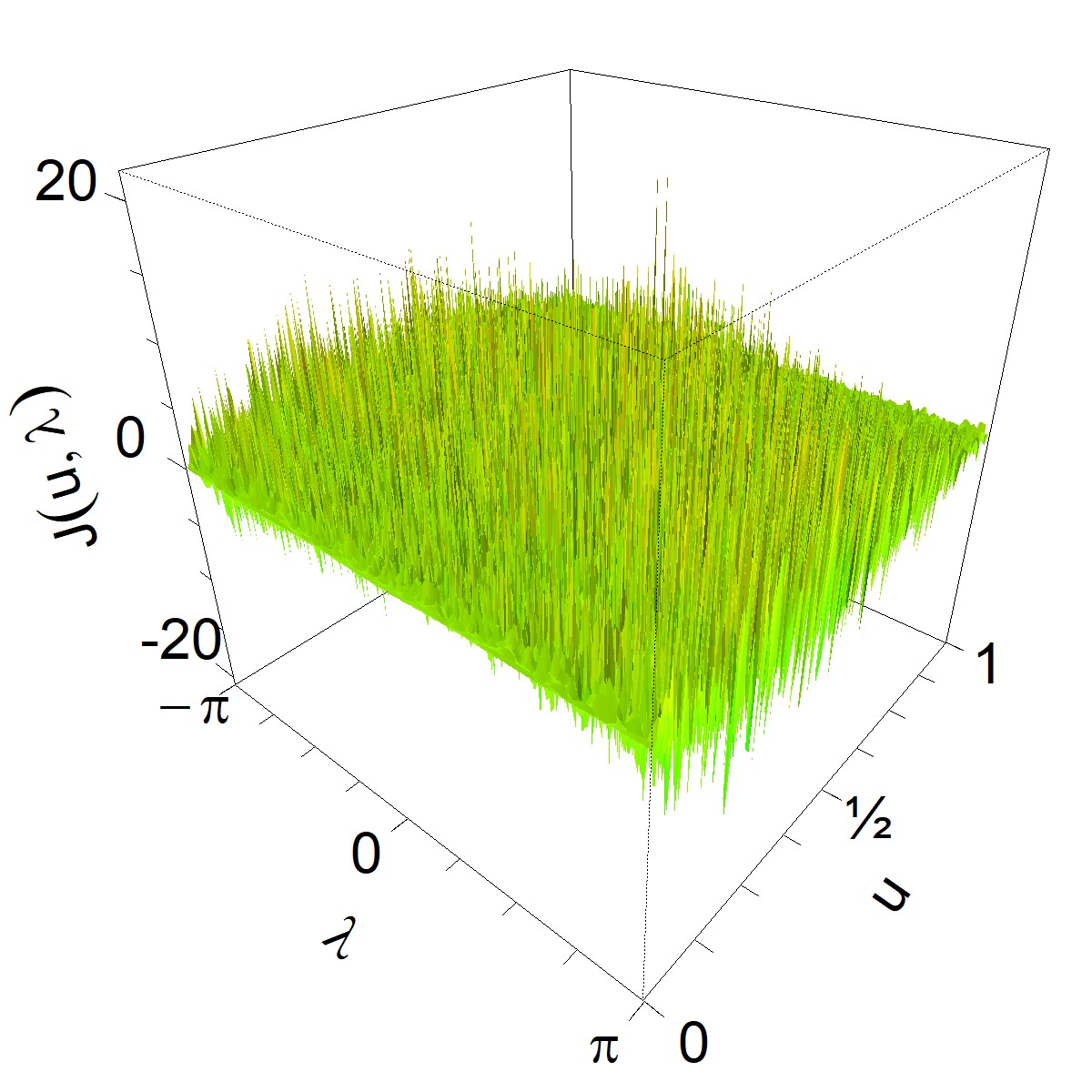}}    
  \put( 5.5,9.3){(b)}                                                       
  \put(11.0,4.7){\includegraphics[width=4.8cm]{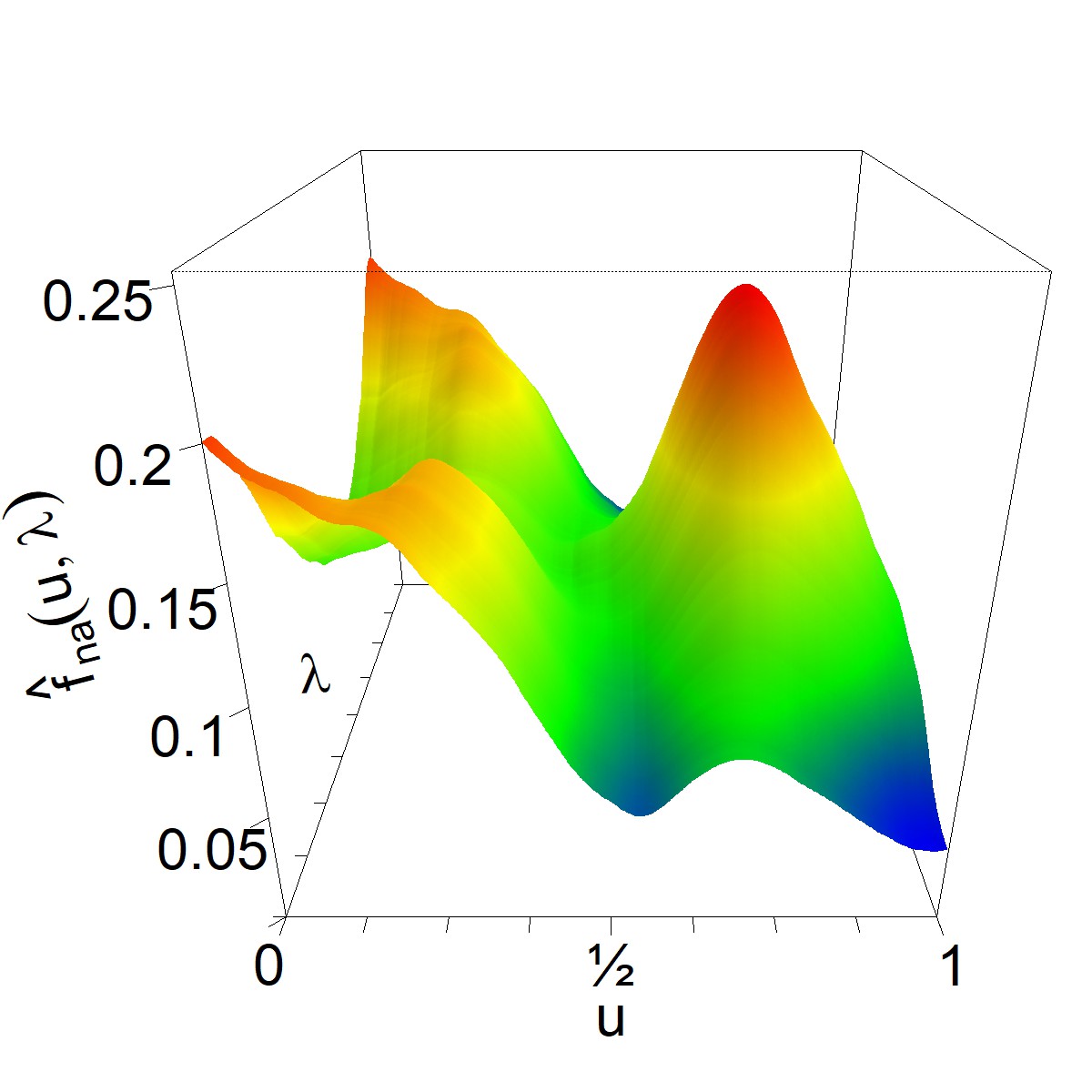}}              
  \put(11.0,9.3){(c)}                                                       
  \put( 0.0,0.0){\includegraphics[width=4.8cm]{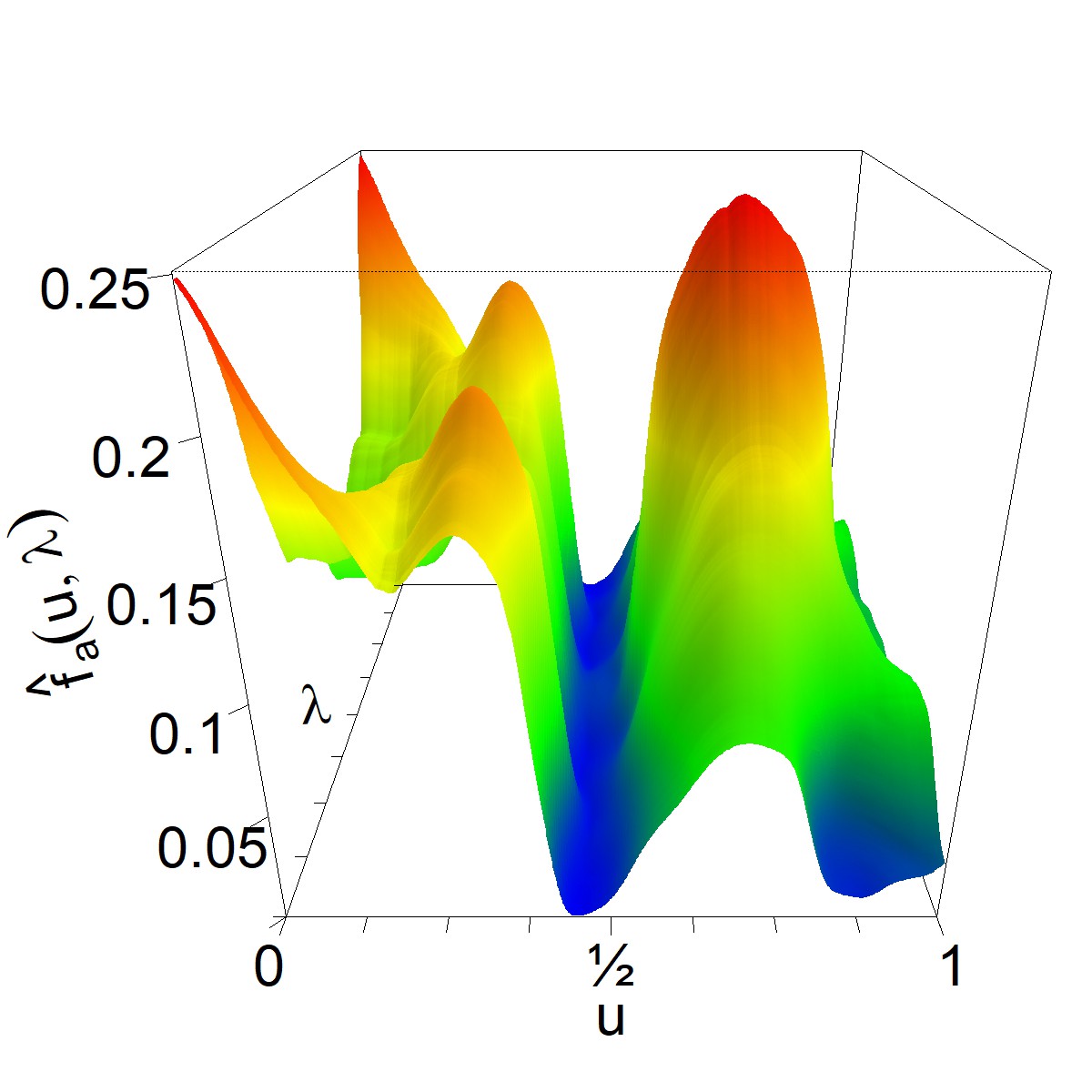}}     
  \put( 0.0,4.5){(d)}                                                       
  \put( 5.5,0.0){\includegraphics[width=4.8cm]{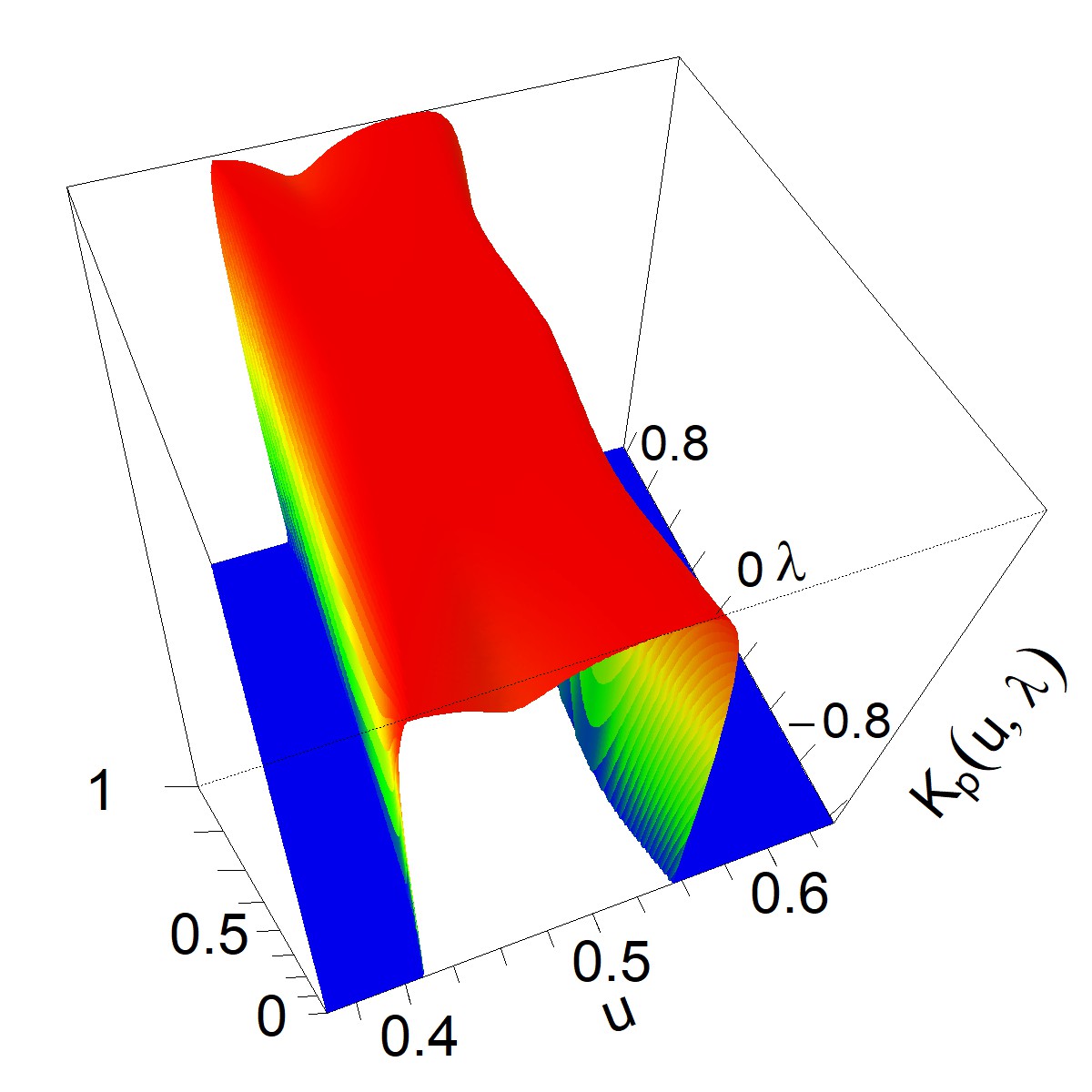}}                  
  \put( 5.5,4.5){(e)}                                                       
  \put(11.0,0.0){\includegraphics[width=4.8cm]{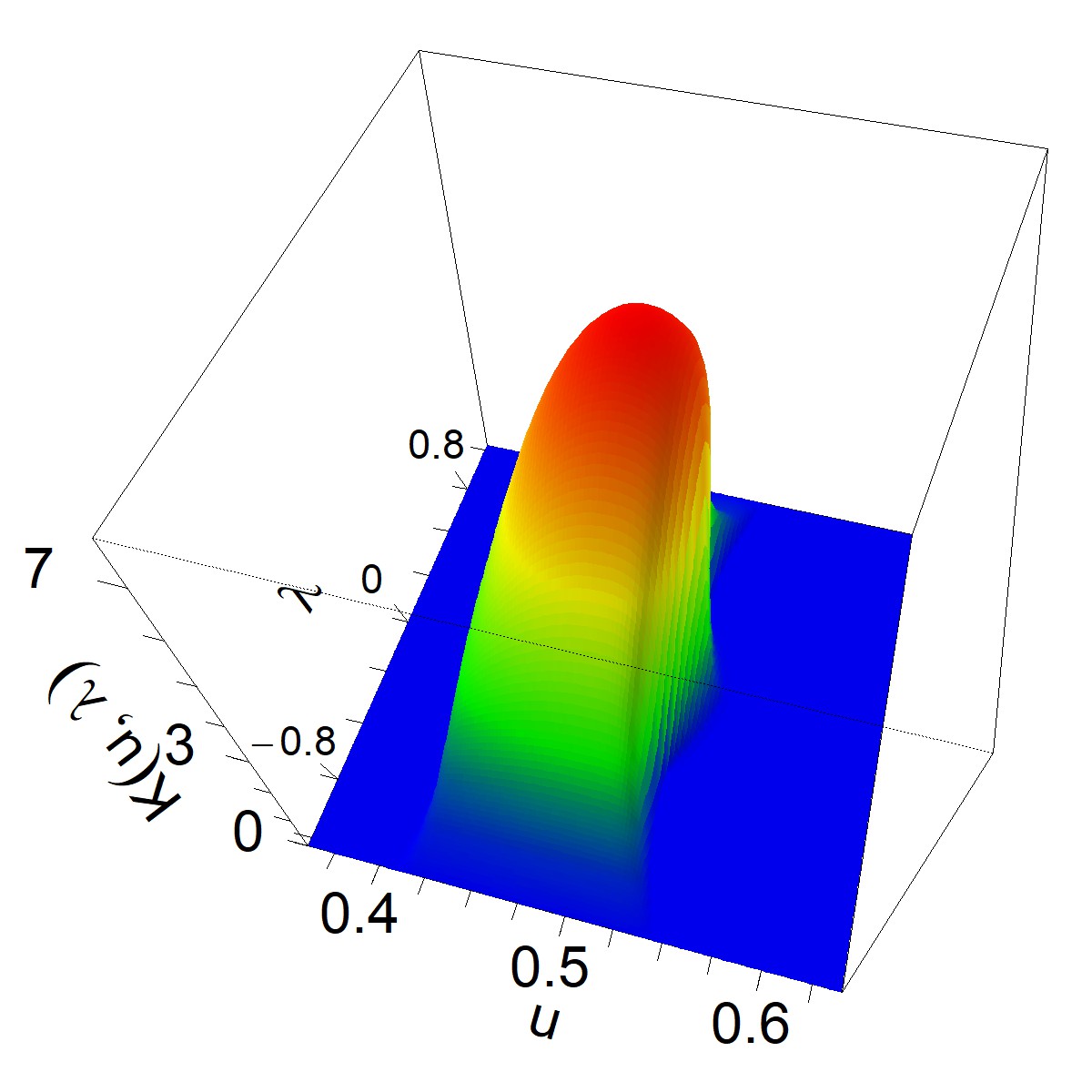}}                 
  \put(11.0,4.5){(f)}                                                       
\end{pspicture}
\caption{Locally stationary process with structural break: (a) true spectrum, (b) pre--periodogram, (c) nonadaptive estimate, (d) adaptive estimate, (e) penalty kernel and (f) smoothing kernel (scaled by a factor $10^5$) for the adaptive estimate at $(u,\lambda)=\big(\tfrac{1}{2},0\big)$.}
\label{fig:masb1}
\end{center}
\end{figure}

We now combine the above two processes. More precisely, we consider a white noise process that at $t_0=0.4\,T$ turns into a moving average process with similar dynamics as in the previous example but shifted in time. Thus we have
\begin{equation} \label{ma2breaklocstat}
Z_{t} =\begin{cases}
\veps_{t}&\text{ if }t \leq t_0\\
\cos\big(2\pi( \tfrac{t}{T}-\tfrac{1}{5})\big)\,\veps_t - \big(\tfrac{t}{T}-\tfrac{1}{5}\big)^2\, \veps_{t-1}&\text{ if }t>t_0
\end{cases}
\end{equation}
with $\veps_t\iid\normal(0,1)$. The corresponding time-varying spectral density is given by
\begin{equation}\label{eq:model1}
 f(u,\lambda)=
\SSS{\frac{\sigma^2}{2 \pi}}\,1_{\{u\leq t_0/T\}}+
g\big(u-\tfrac{1}{5},\lam\big)\,1_{\{u>t_0/T\}},
\end{equation}
where $ g(u,\lam)$ is as in \eqref{ex:locstatdensity}. These type of spectra can for example occur when a signal is constant for a while and then receives a stimulus. This time-varying spectrum is interesting as the peak and the break are both close in distance as well as in level and hence are difficult to resolve.  The adaptive estimate in Figure \ref{fig:masb1}\,(d) was obtained after $k_{\text{final}}=8$ iterations. Both curvature and the break are clearly discernible, and the estimate closely resembles the  true spectrum shown in Figure \ref{fig:masb1}\,(a). Compared with the results for process $X_t$, the flat part of the white noise component is slightly rougher. The corresponding nonadaptive estimated spectrum $\fna$ in Figure~\ref{fig:masb1}\,(c)) evidently suffers from over-smoothing. Figures \ref{fig:masb1}\,(e) and (f) depict the penalty and smoothing kernel for the point $(u,\lambda)=\big(\tfrac{1}{2},0\big)$ which lies in the trough between the break and the peak. Strong penalization in time direction and close to none in frequency direction is visible. This is in accordance with the different slopes in the two directions.

\begin{figure}[h!]
\begin{center}
\includegraphics[width=12cm]{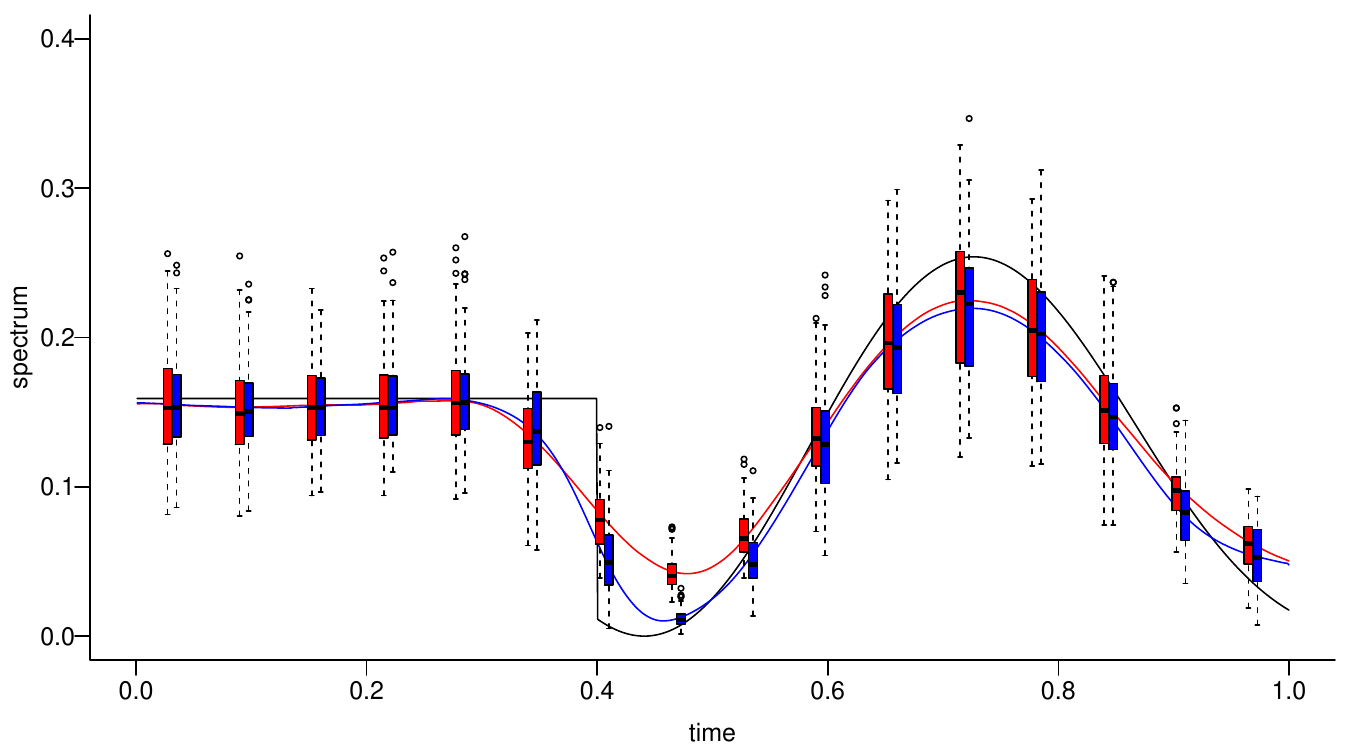}
\caption{Comparison of spectral estimates for process $Z_t$ (locally stationary process with structural break) over 100 repetitions: mean curves and boxplots for the adaptive estimator $\fmax$ (blue) and the oracle kernel estimator $\fnaopt$ (red) over time $u$ for frequency $\lambda=0$; the true spectral density is added in black.}
\label{fig:masb2}
\end{center}
\end{figure}

Compared to the previous case, the simulation study reveals a much better performance of the adaptive estimator than the oracle estimator as shown by the error measures in Table \ref{tab:Errors}. This is also visible in Figure \ref{fig:masb2}. Similar as for process $Y_t$, the adaptive estimator shows less bias in the lower regions and similar results elsewhere. Sightly more variation is observed for the adaptive estimator close to the break, but otherwise it exhibits less variation than the oracle estimator. This is in line with the findings for process $X_t$ and can be explained similarly. Overall, we conclude that the adaptive estimate again captures the features of the true spectrum better than the oracle estimator.

%% file: Sec6-appl.tex
\section{Application to local field potentials}

\begin{figure}[tb]
\begin{center}
  \includegraphics[width=0.6\linewidth]{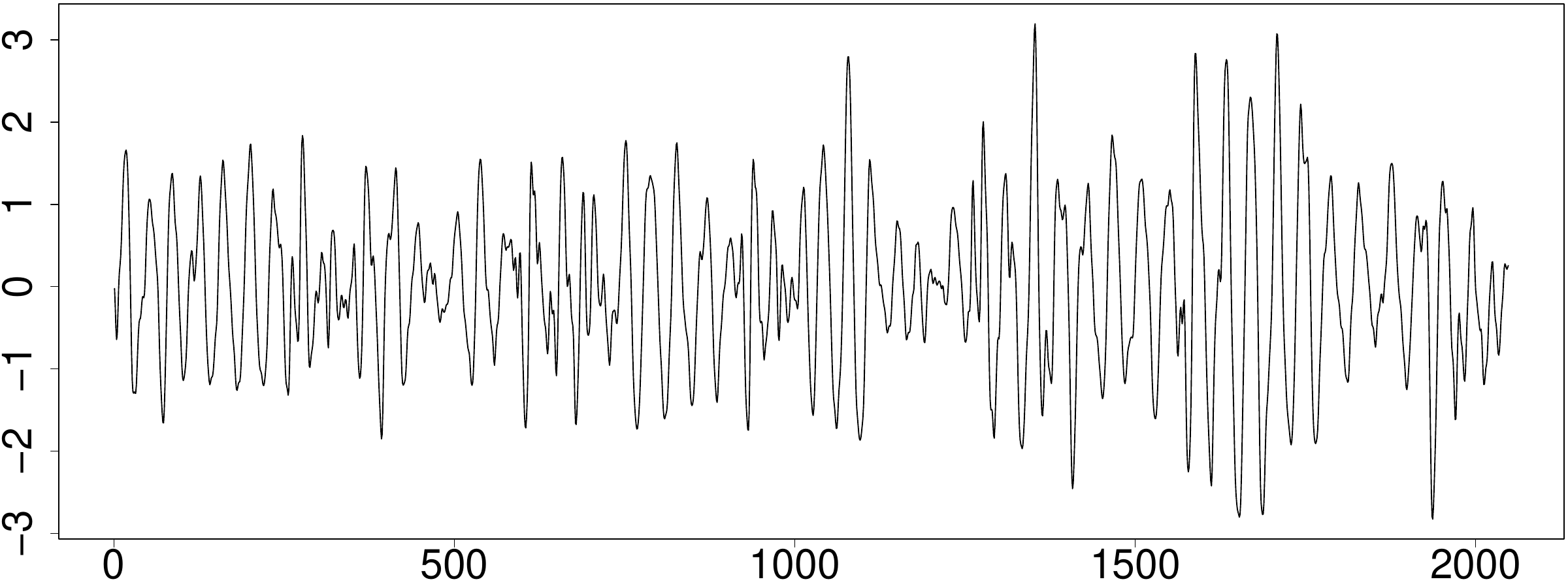}
  \caption{\small{Local field potential recording of length $T=2048$ of the Nucleas accumbens.}}
\label{fig:Nac}
\end{center}
\end{figure}

As an application of our method, we consider local field potential (LFP) recordings of the nucleas accumbens of a male macaque monkey during an associative learning experiment.  These type of data as well as other types of brain data are known to exhibit nonstationary behavior that result in localized signal in the time-varying spectrum, often located in a small frequency band. Existing methods to resolve such signals cannot provide information upon the nature of nonstationarity. Given the ability of our method to resolve localized signal and to find both break structures and smooth patterns we can determine what type of nonstationary behavior characterizes such data. 

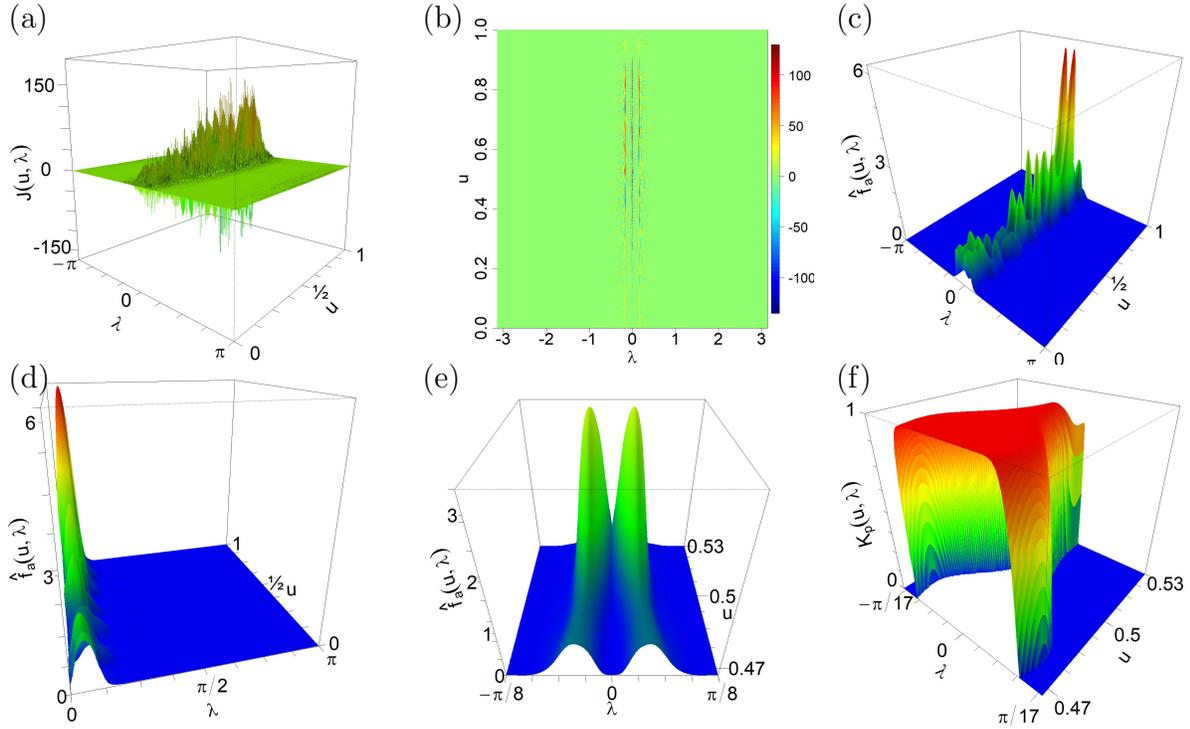
\begin{figure}[tbh]
\begin{center}
\setlength{\unitlength}{1truecm}
\begin{pspicture}(0,0)(16.5,10.2)
  \put( 0.0,4.7){\includegraphics[width=4.8cm]{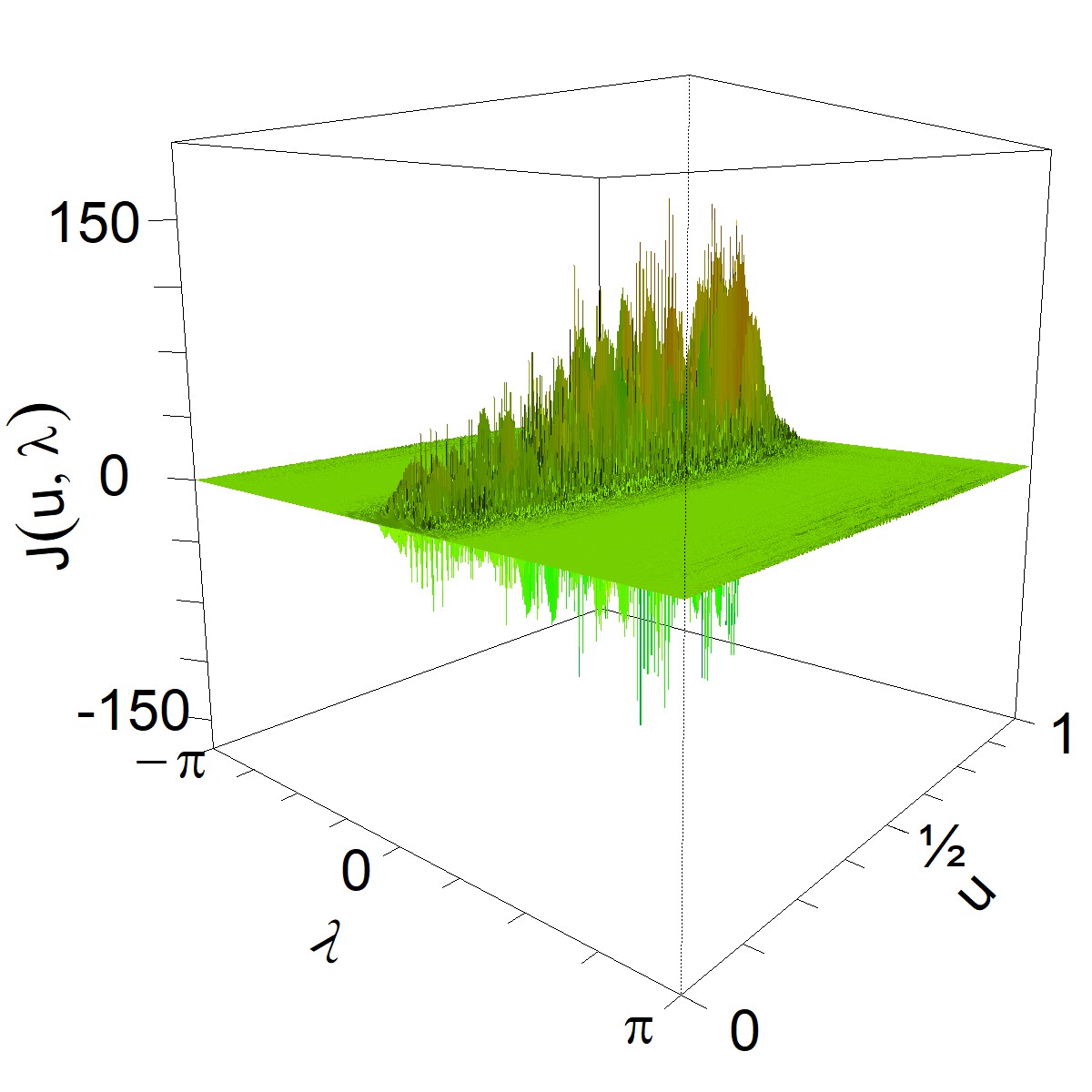}} 
  \put( 0.0,9.3){(a)}                                                       
  \put( 5.9,4.7){\includegraphics[width=4.8cm]{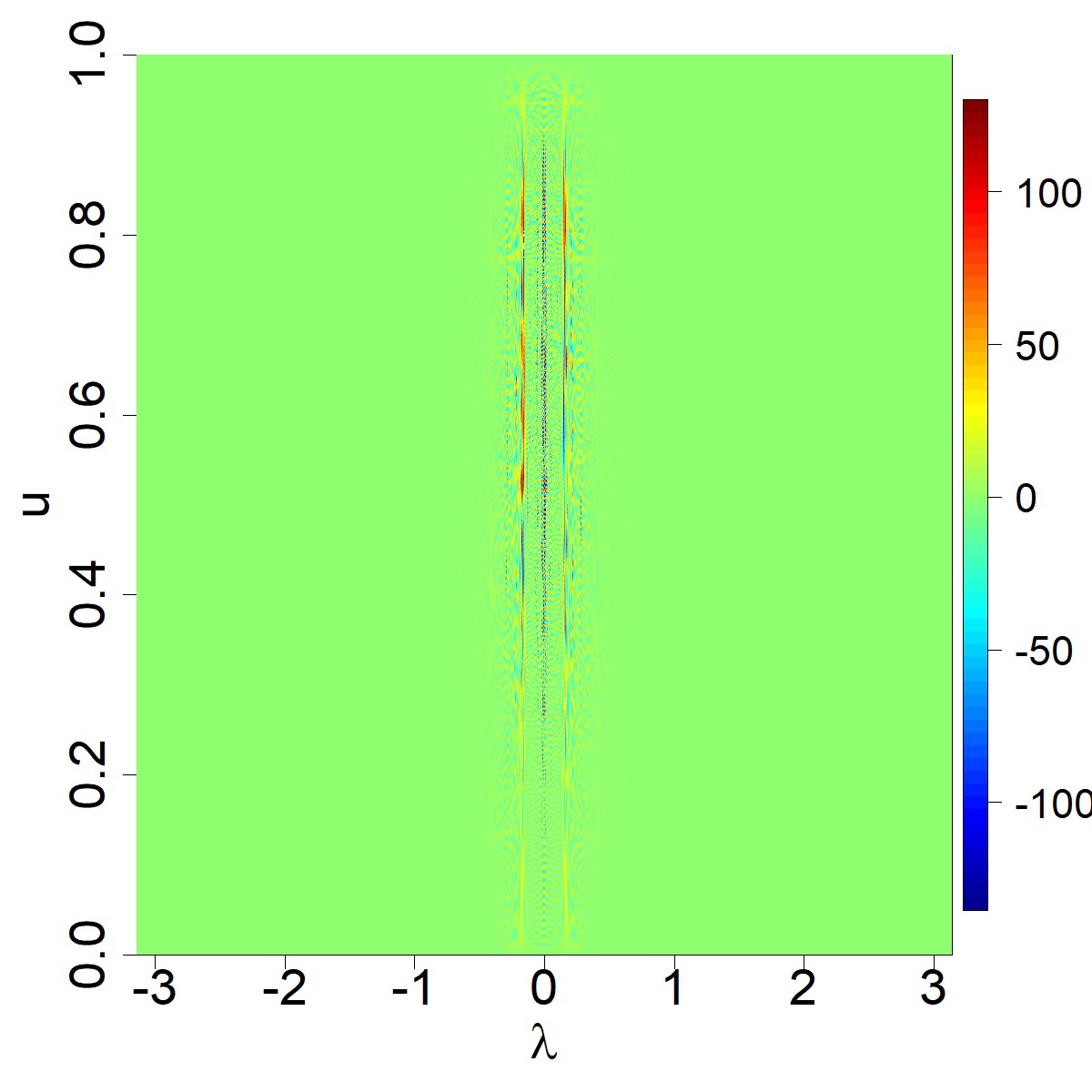}}    
  \put( 5.5,9.3){(b)}                                                       
  \put(11.0,4.7){\includegraphics[width=4.8cm]{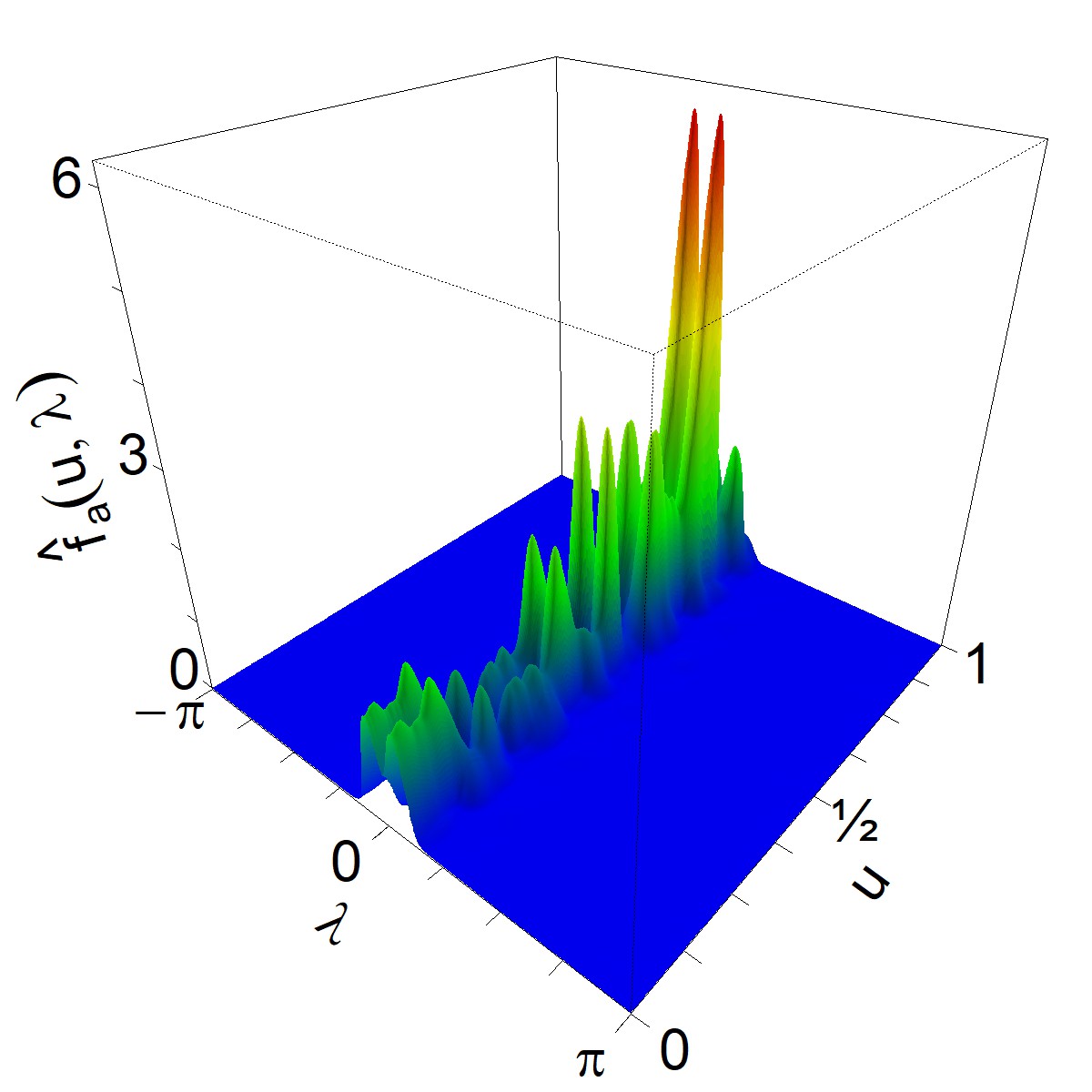}}              
  \put(11.0,9.3){(c)}                                                       
  \put( 0.0,0.0){\includegraphics[width=4.8cm]{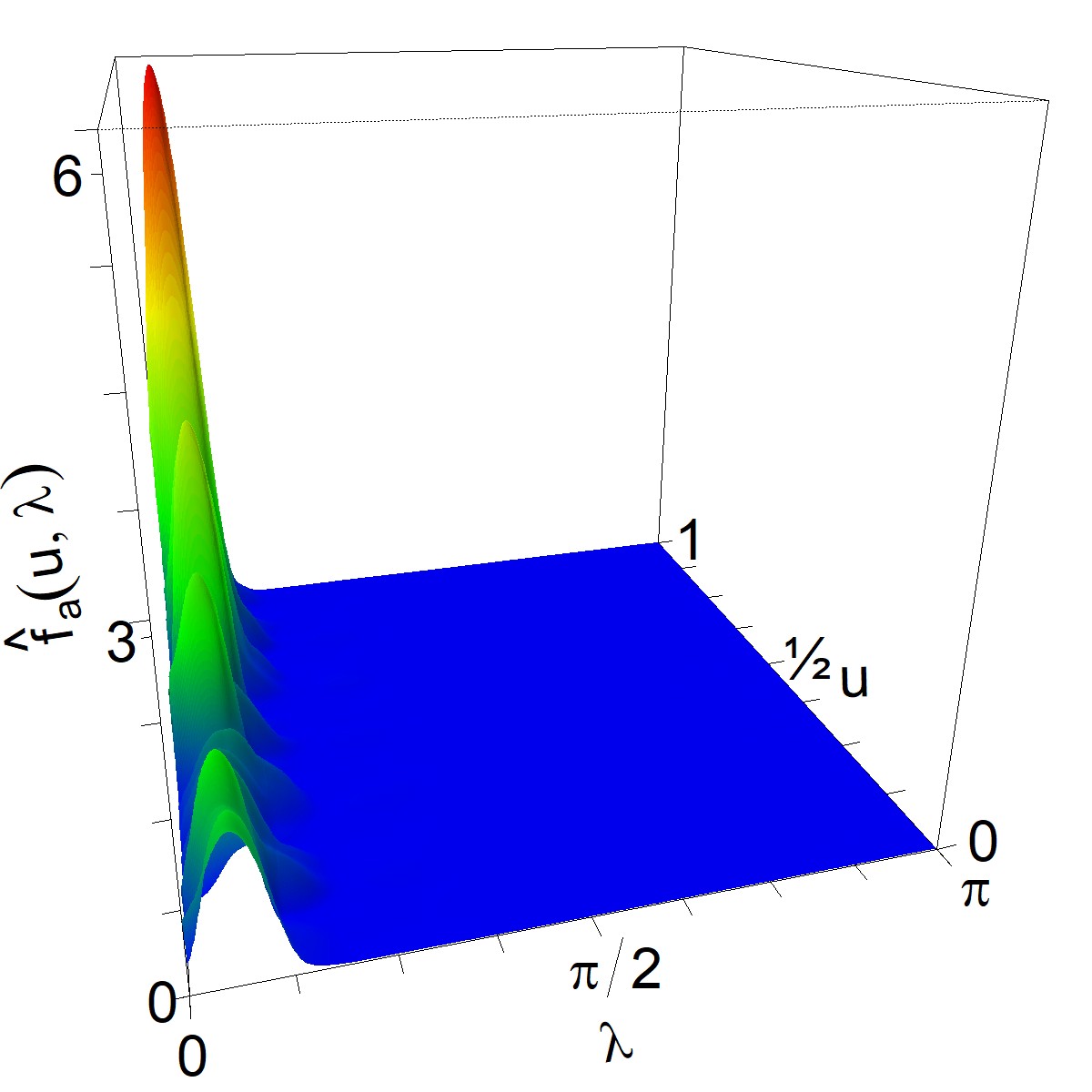}}     
  \put( 0.0,4.5){(d)}                                                       
  \put( 5.5,0.0){\includegraphics[width=4.8cm]{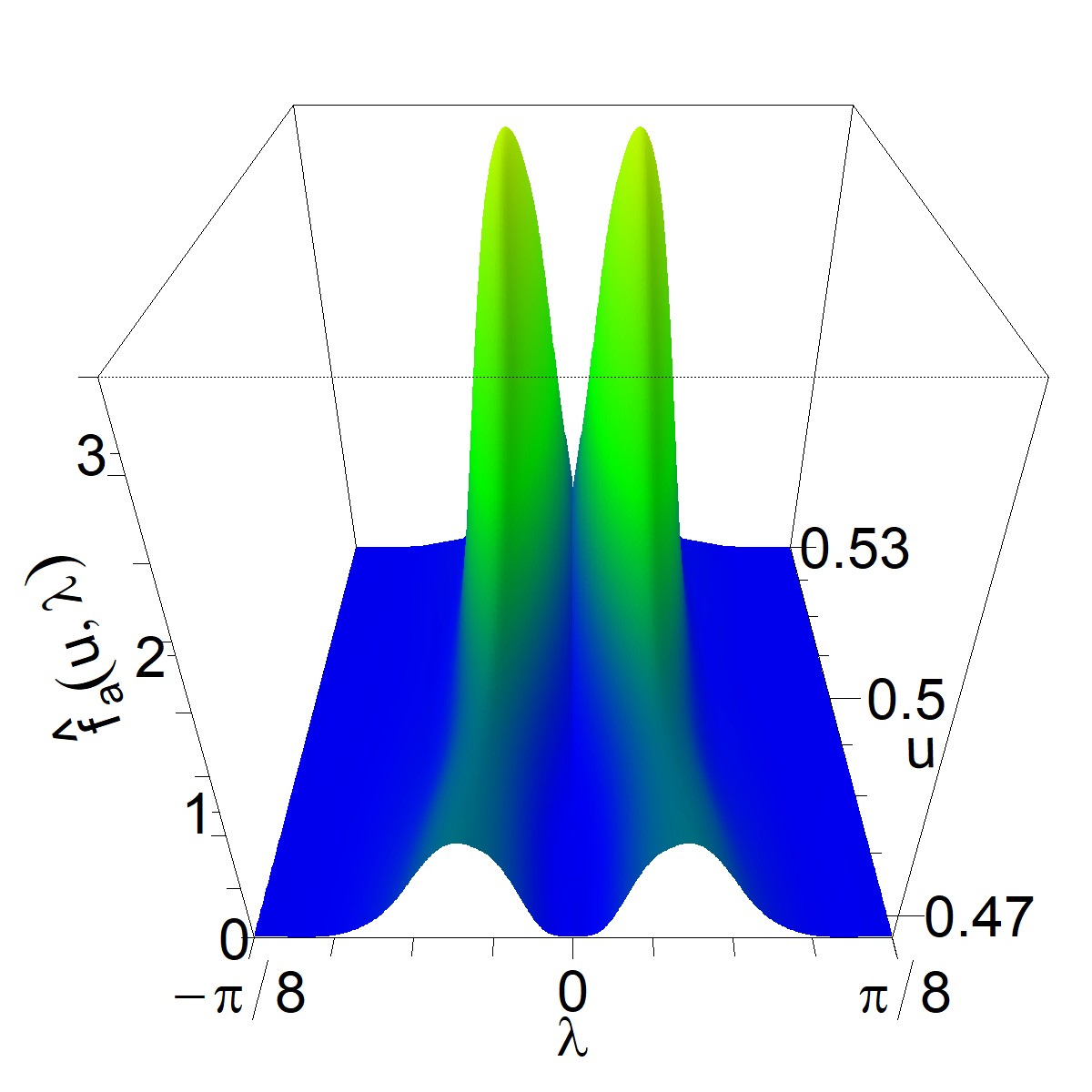}}                  
  \put( 5.5,4.5){(e)}                                                       
  \put(11.0,0.0){\includegraphics[width=4.8cm]{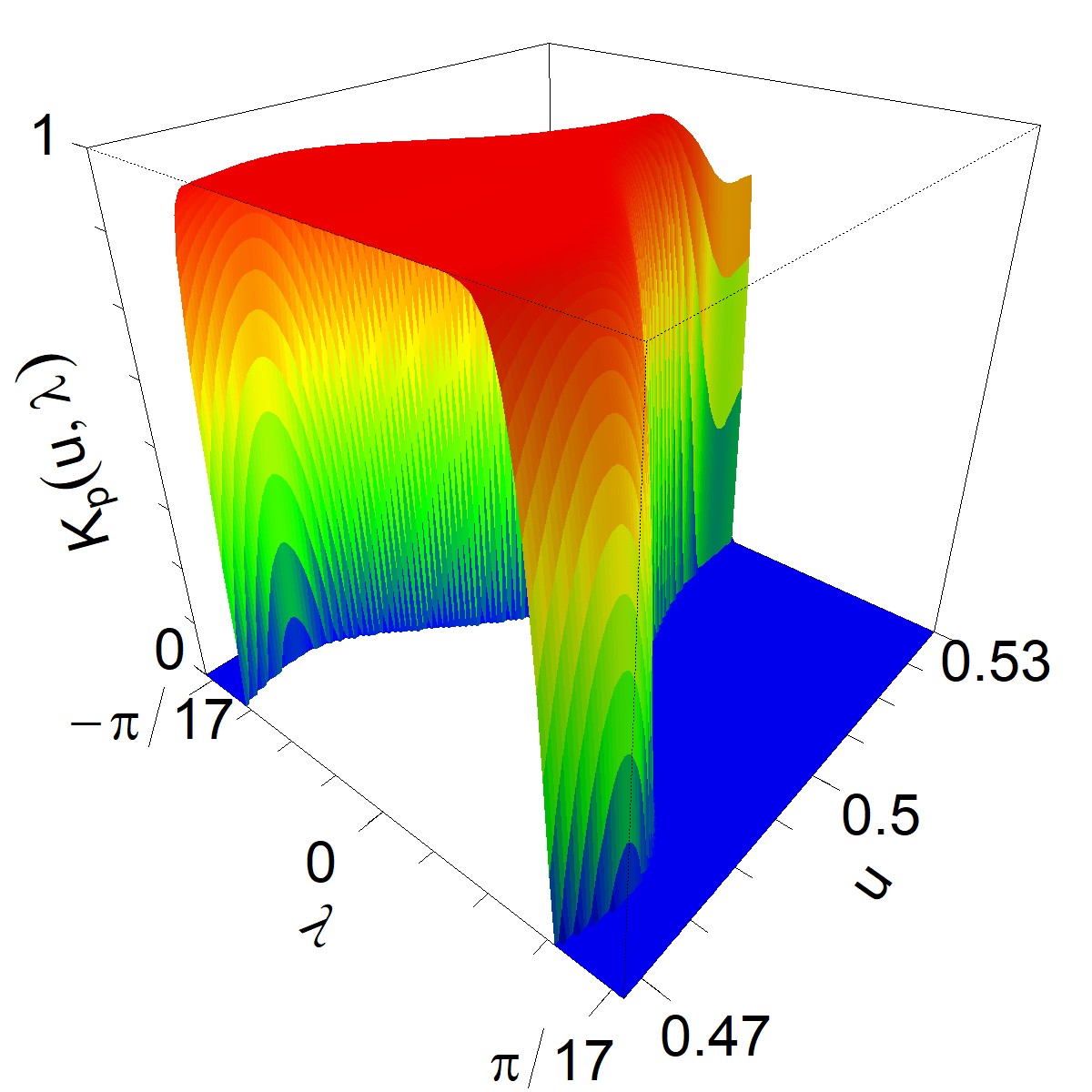}}                 
  \put(11.0,4.5){(f)}  
  \end{pspicture}
\caption{LFP data: (a) perspective and (b) level plot of the pre-periodogram; (c)--(d) adaptive estimate (e) adaptive estimate around $(u,\lambda)=\big(\tfrac{1}{2},0\big)$ and (f) penalty kernel at $(u,\lambda)=\big(\tfrac{1}{2},0\big)$.}
\label{fig:monkey2}
\end{center}
\end{figure}

During each trial of this experiment, the monkey was shown four pictures and then had to select one of four doors. If the monkey made the correct association between picture and door he would receive a reward. In total, the learning experiment, which was conducted at the Neurosurgery Department at the Massachusetts General Hospital, consisted of 675 trials. Figure \ref{fig:Nac} depicts the LFP recordings of one of the trials which is of length $T=2048$. The figure indicates changing dynamics over time.
A first inspection of the pre-periodogram (Fig~\ref{fig:monkey2}\,(a)-(b)) indicates most of the neuronal activity is centered in a small frequency band close to zero. In order to resolve the narrow peak(s), a small bandwidth is required in frequency direction. In particular, it is well-known from the stationary setting the width of the main lobe from the kernel function should be no larger and preferably half the size than the bandwidth of the narrowest peak in the spectrum. Moreover, there is also clear evidence of nonstationary behavior in time direction. Given the length of data, even the relatively conservative default starting bandwidths would out-smooth this behavior. Based on the width of the signal we therefore fix the starting bandwidths to the very low values $\btkkT{0}=\bfkkT{0}=0.025$. The rest of the parameters are set to the default values.


The estimated spectrum is given in Figure \ref{fig:monkey2}\,(c) which was obtained at $k_\text{final}=6$. Signal is estimated in a small frequency band around zero as can be seen more clearly in Figue \ref{fig:monkey2}\,(d). Figures \ref{fig:monkey2}\,(e)-(f) depict the estimated spectrum locally around the point $(u,\lambda)=(\tfrac{1}{2},0)$ together with the penalty kernel obtained for the point $(u,\lambda)=(\tfrac{1}{2},0)$. We observe some definite changes in both width, location and magnitude of the peak over time when looking at the various graphs. Although steep changes occur, they seem smooth in nature. A large spike is visible around the end of the data stretch, possibly indicating the response of the monkey's brain to receiving the award. It is worth noting that despite the low starting bandwidths and the large percentage of cross terms our method overall appears to capture local structures well. 

%% file: Sec7-disc.tex
\section{Conclusion}

In this paper, we introduced a data-driven approach to estimate spectral densities of nonstationary time series, a long time open problem in the analysis of time-varying spectral analysis.  We propose to iteratively determine the optimal shape of the smoothing kernel by determining the maximal neighborhood over which smoothing is justified by the data. This flexibility in shaping the smoothing kernel completely data-adaptively is of particular importance when structural breaks are present. A major problem in high-resolution time-frequency analysis are so-called cross-terms. Our algorithm is specifically designed to control for these cross-terms, which leads to estimates that benefit from the good time-frequency concentration of the pre-periodogram while not suffering from distortions due to cross-terms. A simulation study indicates that our method captures the underlying dynamics of various spectra very accurately and better than \textit{any} kernel estimator with global bandwidth.

One limitation in the application of the algorithm is its complexity as in each step for the shaping of the smoothing kernel every point in the time-frequency plane must be compared with all points in a growing local neighborhood. Although our current implementation makes use of multiprocessor computing, the estimation could become computationally infeasible for very long time series and an extension to high-dimensional time series is --computationally speaking-- not straightforward. This could possibly be improved by the use of graphical processing units. This is left for future work.

%% file: Sec3c-asympt.tex
\section{Asymptotic properties of $\tildefkT(u,\lambda)$} \label{secasym}

The objective of this supplement is to provide some intuition on the distributional properties of the estimator 
\begin{equation}
\label{eq:estimatorad}
\tildefkT(u_r,\lambda_i)
=\tfrac{1}{\tildeNkT(r,i)}\lsum_{(s,j) \in \in B^{(k)}(r,i)}\tildeWkT_{r,i}(s,j)\,\JT\big(u_s,\lambda_j\big)
\end{equation}
with the adapted kernel weights
\begin{equation}
\label{eq:weightsad}
\tildeWkT_{r,i}(s,j)=\Kf\big(\tfrac{\lambda_i-\lambda_j}{\baf^{(k)}}\big)\,
\Kt\big(\tfrac{u_s-u_r}{\bt^{(k)}}\big)\,P^{(k)}_{r,i}(s,j),
\end{equation}
where $\tildeNkT(r,i)=\sum_{s,j}\tildeWkT_{r,i}(s,j)$ and $B^{(k)}(r,i) =\{(s,j):\,|u_s-u_r|<\btT^{(k)},|\lambda_j-\lambda_i|<\bfT^{(k)}\}$ for $k=0,\ldots,k_{\text{max}}$ and where 
\[
\tilde P^{(k)}_{r,i}(s,j)=K_P\Big(\big(\hatfkkT{k-1}(u_r,\lambda_i),\hatfkkT{k-1}(u_s,\lambda_j)\big)\Big).
\]
We start with some necessary background on empirical spectral processes \citep[e.g.][]{Dahlhaus2009b,Dahlhaus2009a}. Generally, the empirical spectral process for arbitrary index functions $\phi$ is defined by 
\begin{align*}
E_T(\phi)= \sqrt{T} \int_{-\pi}^{\pi}\big(F_T(\phi)-F(\phi)\big),
\end{align*}
where 
\begin{align*}
F(\phi) = \int^{1}_{0} \int^{\pi}_{-\pi} \phi(u,\lambda)\,f(u,\lambda)\,du\,d\lambda
\end{align*}
is the generalized spectral measure and
\begin{align*}
F_T(\phi) = \tfrac{1}{T} \lsum_{t=1}^{T} \int^{\pi}_{-\pi} \phi\big(\tfrac{t}{T},\lambda\big)\,J_T(\tfrac{t}{T},\lambda)\,d\lambda
\end{align*}
denotes the corresponding empirical spectral measure. For particular classes of index functions independent of $T$, a functional central limit theorem has been proved \citep[Theorem 2.11]{Dahlhaus2009b}. Additionally, for index functions depending on $T$ a central limit theorem has been derived \citep[Theorem 3.2]{Dahlhaus2009a}.

Many localized statistics for non-stationary time series can be written in terms of the empirical spectral measure. In particular, we obtain the continuous version of the non-adaptive time-varying spectral estimator\citep[(3) of][]{vDE18}, given by
\begin{equation}
\label{eq:fw}
\hatfT(u,\lambda)
=\SSS{\frac{1}{C}}\sum_{s,j}\Kf\Big(\SSS{\frac{\lambda-\lam_j}{\bfT}}\Big)\,
\Kt\Big(\SSS{\frac{u-s/T}{\btT}}\Big)\,\JT\big(\tfrac{s}{T},\lambda_j\big),
\end{equation}
where $\lambda_j=\frac{\pi j}{T}$ for $j=1-T,\ldots,T$ denote the Fourier frequencies and $C=\sum_{s,j}\Kf\big((\lambda-\lam_j)/\bfT\big)\,\Kt\big((u-s/T)/\btT\big)$, paper by considering index functions
\begin{align}
\label{eq:indfnonad}
\phiT(v,\mu) = \tfrac{1}{\btT\,\bfT}\,\Kt \big(\tfrac{u-v}{\btT}\big)\,\Kf\big(\tfrac{\lambda-\mu}{\bfT}\big).
\end{align}
Since the index functions depend on $T$, asymptotic normality of the estimator $F_T(\phiT)$ and its discretized version \eqref{eq:fw}  follows from Theorem 3.2 and Example 4.1 of \citet{Dahlhaus2009a} under the following additional conditions.
\begin{assumption}\label{eq:assum}\mbox{}
\begin{romanlist} 
\item The time-varying spectral density $f(u, \lambda)$ is twice differentiable in $u$ and $\lambda$ with uniformly bounded derivatives.
\item The bandwidths satisfy $\btT,\bfT \to 0$ and $\btT\bfT T\gg\log(T)^2$ as $T \to \infty$, 
\item The kernels $\Kt$ and $\Kf$ are of bounded variation with compact support. 
Moreover, $\int x\,\Kt(x)\,dx = 0$ and $\int\Kt(x)\,dx = 1$ and analogously for $\Kf$.
\end{romanlist}
\end{assumption}
In particular, we find that
\[
\btT\,\bfT\,\var\big(E_T(\phiT)\big)\to 2\pi\,f^2(u,\lam)\,\kappat\,\kappaf.
\]
Furthermore, estimators at different points in the time-frequency plane are asymptotically independent.

For the adaptive estimator in \eqref{eq:estimatorad} similar asymptotic results cannot be derived easily since the final smoothing kernel is iteratively defined and depends on the spectral estimators in previous steps through penalization and the memory step. In the following, we therefore provide at least heuristic arguments that under homogeneity of the spectral density penalization has a negligible effect and hence the estimator remains consistent and asymptotically normal.

More precisely assume that $f(u,\lambda)=f$ for all $u$ and $\lambda$ and define for fixed $u\in[0,1]$ and $\lambda\in[-\pi,\pi]$ the functions
\begin{align}
\psi_{\alpha,\beta} = \phi_{u+\alpha\btT,\lambda+\beta\bfT}^{(T)}
\end{align}
where $\phi^{(T)}_{u,\lambda}$ is defined as above. Then the family of index functions $\fclass_0=\{\psi_{\alpha,\beta}|\alpha,\beta\in[-1,1]\}$ satisfies the conditions of Theorem 2.11 of \citet{Dahlhaus2009b}. Hence the penalty statistic $\Delta(\hatfkT(u,\lambda),\hatfkT(u+\alpha\btT,\mu+\beta\bfT))$ asymptotically has that same distribution as
\[
\SSS{\frac{\btT\,\bfT }{4\pi\,\kappaf\,\kappat\,f^2}}\,
\big(E(\psi_{0,0})-E(\psi_{\alpha,\beta})\big)^2, \tageq
\]
where $E(\psi)$ is a Gaussian process with mean zero and covariances
\begin{align*}
\btT\,\bfT\,\cov\big(E(\psi_{\alpha,\beta}),E(\psi_{\gamma,\delta})\big)
=2\pi&\,f^2\,\int_{-1/2}^{1/2}\int_{-\pi}^{\pi}
\Kt(\alpha-u)\,\Kt(\gamma-u)\,\Kf(\beta-\lambda)\\
&\times\big[\Kf(\delta-\lambda)+\Kf(\delta+\lambda)\big]\,du\,d\lambda+O(\bfT).
\end{align*}
The expression shows that under the assumption of homogeneity of the time-varying spectrum over the local neighborhood about the point $(u,\lambda)$ the distribution of the penalty statistic does not depend on the bandwidth or the sample size but through a term of order $O(\bfT)$. Moreover, the strong positive correlation of the Gaussian process $E(\psi)$ leads to at most weak penalization towards the borders of the local neighborhood yielding a total smoothing kernel that differs only slightly from the non-adaptive smoothing kernel. Finally, since $E(\psi_{0,0})$ and $E(\psi_{\alpha,\beta})$ are positively correlated, the variance of their difference can be bounded by $2\,\var\big(E(\psi_{0,0})\big)$ uniformly for all $\alpha,\beta\in[-1,1]$, which justifies the use of the $\chi^2_1$-distribution for determining the cut-off point of the penalty kernel.

We note that the same covariance structure can be derived from Theorem 3.2 of \citet{Dahlhaus2009a} by considering the index functions $\phi_{u+\alpha\btT,\lambda+\beta\bfT}^{(T)}$ directly, that is, taking  their dependence on $T$ into account in the asymptotics. However, the result is weaker insofar it does not yield convergence over the whole local neighborhood defined by $\alpha,\beta\in[-1,1]$ simultaneously. Although the above arguments based on fixed index functions indicate that this result could be strengthened, a derivation of a functional central limit theorem in this setting is beyond the scope of this paper. \\

Summarizing we find that under the assumption of homogeneity penalization does only modify the shape of the smoothing kernel even if applied iteratively multiple times but will keep the rates approximately the same. In contrast, in case of a non-constant spectral density, the penalty statistic depends quadratically on the difference in levels which leads to more severe penalization as bandwidths in time and frequency direction increase. Accordingly, the resulting smoothing kernel will have in general a smaller support corresponding to a smaller bandwidth than the one actually imposed. Nevertheless, in the setting of locally stationary processes this effect will disappear asymptotically since the level of local homogeneity increases as long as the bandwidths used in the iteration satisfy the conditions in Assumption \ref{eq:assum}. In other words, the adaptive estimator remains consistent with rate $\sqrt{T\btkkT{k_\text{max}}\bfkkT{k_\text{max}}}$ since its adaptiveness only shows in finite samples. This is even true when the dynamics of the process exhibit structural breaks and thus should be described by a piecewise locally stationary process. In that case, penalization will be strong in the local neighborhood of a break leading to asymmetric smoothing kernels that seem to be cut off. Again, since the local neighborhoods (in the rescaled time-frequency plane) are shrinking for increasing sample size, the effect will disappear but for the points along the breaks where the time-varying spectral density is not well-defined. Examples of such processes with structural breaks are discussed in Section 4 of the main paper, where we illustrate the final sample behavior of the adaptive estimator by simulations.
